\documentclass[11pt,a4paper]{article}
\pdfoutput=1

\usepackage{amsmath,amssymb,tikz,hyperref,empheq}
\usepackage{graphicx}
\usepackage{cite}

 \allowdisplaybreaks
\def\be{\begin{equation}}
\def\ee{\end{equation}}
\def\ba#1\ea{\begin{align}#1\end{align}}
\def\bg#1\eg{\begin{gather}#1\end{gather}}
\def\bm#1\em{\begin{multline}#1\end{multline}}
\def\bmd#1\emd{\begin{multlined}#1\end{multlined}}

\def\a{\alpha}

\def\d{\delta}
\def\D{\Delta}
\def\e{\epsilon}   

\def\g{\gamma}
\def\G{\Gamma}

\def\k{\kappa}
\def\l{\lambda}

\def\r{\rho}

\def\la{\label}

\def\nn{\nonumber}

\def\({\left(}
\def\){\right)}
\def\[{\left[}
\def\]{\right]}

\def\cO{{\mathcal O}}

\def \be {\begin{equation}}
\def \ee {\end{equation}}
\def \ba {\begin{array}}
\def \ea {\end{array}}
\def \bea{\begin{eqnarray}}
\def \eea{\end{eqnarray}}
\def \nn {\nonumber}

\def \a {\alpha}

\def \g {\gamma}

\def \G {\Gamma}
\def \d {\delta}
\def \D {\Delta}
\def \e {\epsilon}

\def \k {\kappa}
\def \l {\lambda}

\def \r {\rho}

\def \vy {\vec{y}}

\def \la {\leftarrow}
\def \ra {\rightarrow}

\def \ha  {{\hat a}}

\def\bea{\begin{eqnarray}}
\def\eea{\end{eqnarray}}
\newcommand{\eq}[1]{(\ref{#1})}
\newcommand{\bit}{\begin{itemize}}  \newcommand{\eit}{\end{itemize}}
\newcommand{\ben}{\begin{enumerate}}  \newcommand{\een}{\end{enumerate}}

\def\la{\langle}
\def\ra{\rangle}

\def\cG{{\cal G}}  
  
  \def\cO{{\cal O}}

\long\def\symbolfootnote[#1]#2{\begingroup%
\def\thefootnote{\fnsymbol{footnote}}\footnote[#1]{#2}\endgroup}

\newcommand{\nthu}{{\it Department of Physics, National Tsing-Hua
  University,
  Hsinchu 30013, Taiwan}}

\newcommand{\ctc}{{\it
Center of Theory and Computation, 
National Tsing-Hua University, Hsinchu 30013, Taiwan}}

\newcommand{\sysu}{{\it School of Physics and Astronomy, Sun Yat-Sen University, 2 Daxue Road, Zhuhai 519082, China}}

\begin{document}
\thispagestyle{empty}
\begin{center}

~\vspace{20pt}

  {\Large\bf  Conformal Boundary Condition and
Massive Gravitons  in AdS/BCFT}

\vspace{25pt}

Chong-Sun Chu ${}^{2,3}$\symbolfootnote[1]{Email:~\sf
  cschu@phys.nthu.edu.tw}
, Rong-Xin Miao ${}^1$\symbolfootnote[2]{Email:~\sf
  miaorx@mail.sysu.edu.cn}

\vspace{10pt}${{}^{1}}$\sysu \footnote{All the Institutes of authors
  contribute equally to this work, the order of Institutes is adjusted
  for the assessment policy of SYSU.}

\vspace{5pt}${{}^{2}}$\ctc

\vspace{5pt}${{}^{3}}$\nthu

\vspace{1cm}

\begin{abstract}
  According to Witten [1], 
  the conformal boundary condition of gravity, which
  specifies the conformal geometry of the boundary and the trace of
  the extrinsic curvature, is elliptic and leads to
  well-defined perturbation theory
  of gravity about any classical solution.
  The conformal boundary condition was previously considered in
  \cite{Miao:2017gyt,Chu:2017aab}
  in the context of AdS/BCFT, wherein the equation of motion of the
  end-of-the-world was derived and emphasized. In this paper,
  we investigate further other consequences of the conformal boundary condition
  in AdS/BCFT.
  We derive the boundary central charges of the holographic Weyl anomaly
  and show that 
they are exactly the same for conformal boundary condition and Dirichlet
  boundary condition.
  We  analysis the metric perturbation
  with conformal boundary condition (CBC), Dirichlet boundary condition
  (DBC)
  and Neumann boundary condition (NBC)
  imposed on the end-of-the-world brane
  and show that they
  admit an interpretation as the fluctuation
  of the extrinsic curvature (case of CBC and DBC) and the
  induced metric (case of NBC) of $Q$ respectively.
  In all cases,
  the fluctuation modes are massive,
  which
  are closely relevant to the massive island formation in the literature.
  Our results reveal that there are non-trivial gravitational dynamics
  from extrinsic curvatures on the conformal and Dirichlet branes,
  which may have
  interesting applications to the island.  
  We also discuss, in passing,
  the localization of gravitons in brane world theory.
  We find that, contrary to NBC,
  the graviton for CBC/DBC is located on the brane with non-positive
  tension instead of non-negative tension. 
  
\end{abstract}

\end{center}

\newpage
\setcounter{footnote}{0}
\setcounter{page}{1}

\tableofcontents

\section{Introduction}

Double holography has drawn much attention recently,
which plays an important role in recovering Page curve of Hawking
radiation \cite{Penington:2019npb,Almheiri:2019psf,Almheiri:2019hni}.
See also
\cite{Rozali:2019day, Chen:2019uhq,Almheiri:2019psy,Kusuki:2019hcg,
   Balasubramanian:2020hfs,Sully:2020pza,Geng:2020qvw,Chen:2020uac,
  Dong:2020uxp,Arias:2019zug,Arias:2019pzy,Geng:2020kxh,Ling:2020laa,
   Geng:2020fxl,Kawabata:2021hac,Bhattacharya:2021jrn,Kawabata:2021vyo,
   Geng:2021hlu,Krishnan:2020fer,Deng:2020ent,Chu:2021gdb,Neuenfeld:2021wbl,
  Neuenfeld:2021bsb,Chen:2020hmv,Ghosh:2021axl,Omiya:2021olc}
for related works. Double holography is a generalization of the
AdS/CFT correspondence
\cite{Maldacena:1997re,Gubser:1998bc,Witten:1998qj} and is closely
related to brane world holography
\cite{Randall:1999ee,Randall:1999vf,Karch:2000ct} and AdS/BCFT
\cite{Takayanagi:2011zk,Fujita:2011fp,Nozaki:2012qd},
\cite{Miao:2017gyt,Chu:2017aab,Miao:2018qkc}. Here
BCFT means a conformal field theory defined on a manifold with a
boundary, where suitable boundary conditions (BC) are imposed
\cite{Cardy:2004hm, McAvity:1993ue}. Recently, a novel doubly
holographic model called wedge holography has been proposed
\cite{Akal:2020wfl,Bousso:2020kmy}.
For one novel class of solutions, it
has been shown that wedge holography is
equivalent to AdS/CFT with Einstein gravity \cite{Miao:2020oey}.
Generally, wedge holography can
be obtained as a special limit of AdS/BCFT
with edge modes living on
the corner of the wedge \cite{Akal:2020wfl}.
Generalizing wedge holography to
codim-$m$ defects, \cite{Miao:2021ual} 
proposes the so-called cone
holography.

Due to the importance of AdS/BCFT itself and the prominent role it
plays in various other dualities as discussed above,
any deeper understanding of AdS/BCFT will be quite interesting. 
At the level of classical gravity, the action for AdS/BCFT is given by
\begin{equation}\label{action}
I=\int_{N}d^{d+1}x \sqrt{|g|}( R-2\Lambda ) +2\int_Q d^dy\sqrt{|h|} (K -T)
\end{equation}
where $K$ is the extrinsic curvature, $T$ is the tension of
end-of-the-world brane $Q$ and $h_{ij}$ is the induced metric on $Q$.
Taking the variations and focusing on the boundary terms, we have
\begin{equation}\label{variation}
\d I = -\int_Q d^dy \sqrt{|h|}  \left(K^{ij} -(K-T)h^{ij} \right) \delta h_{ij} =0
\end{equation}
in order to have a well-defined action principle (\ref{variation}).
The original proposal of  Takayanagi
\cite{Takayanagi:2011zk} is to take
on $Q$ the Neumann boundary condition (NBC) 
\begin{equation}\label{NBC}
\text{NBC}: \ \left(K^{ij} -(K-T)h^{ij} \right)|_Q=0.
\end{equation}
The Neumann boundary condition 
imposes conditions on the end-of-the-world
brane $Q$ \cite{Takayanagi:2011zk,Fujita:2011fp,Nozaki:2012qd}
as well as the bulk Einstein metric \cite{Miao:2017aba}.
In addition, it is  possible to impose alternative boundary conditions
for AdS/BCFT. For example, we have
the conformal boundary condition (CBC) \cite{Miao:2017gyt,Chu:2017aab},
which fixes the conformal geometry
of the boundary and the trace of the extrinsic curvature 
\begin{subequations}\label{CBC}
  \begin{empheq}[left=\text{CBC}:\empheqlbrace]{align}
&
K=\frac{d}{d-1} T, \label{CBCa}\\
&  \delta h_{ij}|_Q= 2\sigma(y) h_{ij}|_Q, \label{CBCb}
  \end{empheq}
\end{subequations}
where $\sigma(y)$ is an arbitrary conformal factor. One can also impose the 
Dirichlet boundary condition (DBC) \cite{Miao:2018qkc}
\begin{equation}\label{DBC}
\text{DBC}: \ \delta h_{ij}|_Q=0.
\end{equation}
All of these boundary conditions define consistent theory of AdS/BCFT. 
See also \cite{Anderson:2006lqb,Anderson:2007jpe,Anderson:2010ph,York:1972sj,Papadimitriou:2005ii} for some early works on the boundary condition of the gravity. 

To see that the CBC works, it is instructive to  rewrite (\ref{variation}) as 
\begin{equation}\label{variation1}
  \int_Q d^dy \sqrt{|h|}  \big{[ } (T-\frac{d-1}{d} K) \delta h
    + \bar{K}^{ij}  \delta h_{ij} \big{]}=0,
\end{equation}
where $\delta h=h^{ij}\delta h_{ij}$ is the trace of metric variations
and $\bar{K}^{ij}:=K^{ij}-\frac{K}{d} h^{ij}$ is the traceless part
of the extrinsic curvature.  Since $\bar{K}^{ij} \delta h_{ij}=2\sigma
\bar{K}^{ij}h_{ij}=0$, it is clear that the CBC (\ref{CBC}) makes
vanish the action variation (\ref{variation1}). It should be mentioned
that the CBC (\ref{CBC}) has been partially discussed in
\cite{Miao:2017gyt,Chu:2017aab}, which was called mixed BC
there. However, \cite{Miao:2017gyt,Chu:2017aab}
have mainly focused on the
constraint \eq{CBCa}
without paying much attention to the
second condition \eq{CBCb}.
In this paper, we will
study carefully the complete CBC in order
to gain further insights on AdS/BCFT with
CBC.

CBC is not just a possible boundary condition for AdS/BCFT, it is actually
a very interesting class of boundary condition for a good reason.
At the quantum level, one hopes the boundary condition of gravity to be
elliptic so that
it leads to a well-defined perturbation theory of `quantum gravity'
\cite{Witten:2018lgb}. According to Witten \cite{Witten:2018lgb},
in general DBC is not
elliptic and does not lead to a well-defined perturbation
theory. It is better-behaved if the extrinsic curvature of the
boundary is
positive- or negative-definite. This additional
condition indeed plays an important role in AdS/BCFT with DBC, which
helps
to select the correct solutions with positive brane tensions
\cite{Miao:2018qkc}.  On the other hand, CBC is
always
elliptic and leads to
a well-defined perturbation theory \cite{Witten:2018lgb}. Thus it is
interesting to
consider AdS/BCFT with CBC and investigate the properties of the gravitational
perturbations in this context. This is the main
motivation of this paper. 

The analysis of AdS/BCFT with CBC is however a little more subtle.
As originally suggested in \cite{Nozaki:2012qd}
and fully developed in \cite{Miao:2017aba}, it has been
shown that while the traditional construction of the bulk metric based on the
Fefferman-Graham (FG)
expansion does not work for AdS/BCFT due to the existence of junction,
a construction of the bulk
metric based on a perturbative expansion in the extrinsic curvature works
fine. This method has been applied for the case of NBC \cite{Miao:2017aba}
and DBC \cite{Miao:2018qkc} with the bulk metric constructed
correspondingly, leading to well-defined AdS/BCFT with NBC and DBC.
However, it has been observed that the employment of CBC (\ref{CBC}) does
not fix all the integral constants
in the bulk metric solution of the Einstein equations
 \footnote{See also discussions at the
   end of \cite{Miao:2018qkc}, which
   suggests that ``CBC is more subtle,
   which is less restrictive than DBC and NBC."}. This is puzzling and appears
 to be an obstacle to the construction of a well-defined
 theory of AdS/BCFT based on CBC. In this paper, we
 resolve this problem
 and show that CBC also fixes the bulk metric completely and give rises to
 a well-defined AdS/BCFT.
 We observe that the metric ansatz employed in \cite{Miao:2018qkc,Miao:2017aba}
 admits a non-vanishing extrinsic curvature $k_{ij}$
 for the boundary metric of BCFT.
 For simplicity,  we have considered a constant $k_{ij}$ and that was
 sufficient for the analysis there since the NBC and DBC were already
 nontrivial at the considered linear perturbative order.  
 However, as the CBC involves higher power of derivatives in the boundary
 metric, it becomes trivial at the linear perturbative order and this is why
 the CBC appears to be less restrictive. In this paper, we construct the metric
 ansatz with non-constant $k_{ij}$ and show that the CBC does fix the
 bulk metric completely as the case of NBC and DBC. As a result, all the
 boundary central charges of the Weyl anomaly are also determined.
 
 
 Let us summarize the main results of this paper. We investigate
 AdS/BCFT with CBC in this paper.  We resolve a
 subtly related to the application of the CBC to the bulk metric ansatz 
that has been shown to work well previously for NBC and DBC.
We fix the central charges of boundary Weyl anomaly for AdS/BCFT with CBC.
It is found that
 the central charge, which is related to the norm of displacement
 operator, is the exactly the same for CBC and DBC. Although the
 central charges are same, the locations of end-of-the-world branes are
 different for these two
 AdS/BCFT.  We  study the dynamics
 of the metric perturbations
  with CBC, DBC or NBC
imposed on the end-of-the-world branes.
 At the linear order,
 the perturbations obey EOM of massive gravity
 with a mass square spectrum that is discrete and
positive.
The perturbation modes are obtained by imposing a gauge
fixing condition. They can be quantized covariantly by utilizing
the BRST method with the inclusion of ghost fields.
As shown in \cite{Witten:2018lgb},
the resulting kinetic operators in the BRST invariant action
are elliptic.
It is in this sense that these fluctuations
lead to a well-defined perturbative theory of gravity at the quadratic order. 
As application, we
also briefly discuss the potential implications of the fluctuations,
 such as the two-point function and the relation to massive islands
 \cite{Geng:2020qvw,Geng:2021hlu}.

 We emphasis that in AdS/BCFT, the boundary condition is defined with
 a one side end-of-the-world brane, i.e. there is nothing beyond it. This is
 in contrast with the consideration in the Randall-Sundrum (RS) brane world
 theory where the RS brane is situated in a bulk spacetime and a junction
 condition is imposed instead. To get to a BCFT, one needs to perform an
 orbifolding to get to a one-sided setup. Apart from giving an
 extra factor of 2 in the equation of motion for $Q$ in the case of
 NBC and CBC, it also has implication on the localization of gravitons
 due to the rise of a volcano potential localized on the RS brane.
 Although it is not directly related to the study of AdS/BCFT, the set up is
 similar and we have performed a study of the problem of
 localization of gravitons in the brane world setup.
 Interestingly we find
 that, for NBC, the graviton is located on the brane with non-negative
 tension; while for CBC, the graviton is located on the brane with
 non-positive tension. This analysis is included in the appendix.

The paper is organized as follows.  In section 2, we formulate
AdS/BCFT with CBC and resolve a subtlety. We study the linear
perturbations around an AdS background and find that the more general
metric ansatz can fix all of the integral constants and yield a
well-defined CBC. In section 3, we study the perturbations around a
black string and show that CBC is well-defined on this background. In
section 4, we discuss the second-order perturbations and determine
all the central charges of 4d holographic BCFT with  CBC.
In section 5, we explore the dynamics of metric fluctuations
on Dirichlet and conformal branes. 
In section 6, we make
some physical discussions in relation to our results.
In particular, we discuss
the relation of our results to the massive island 
and the
consistency of
the various boundary conditions. 
Finally, we conclude with some open
questions in section 7. A number of appendices is included which
extended the results of the main text of the paper. In appendix A, we solve
the Einstein equation up to second order in small extrinsic curvature of 
 the field-theory boundary.
In appendix B, we derive the massive spectrum for a massive bulk scalar field.
In appendix C, localization of gravitons on the AdS brane is discussed for
various kinds of BC imposed on the brane.

\section{AdS/BCFT with CBC}

\begin{figure}[t]
\centering
\includegraphics[width=10cm]{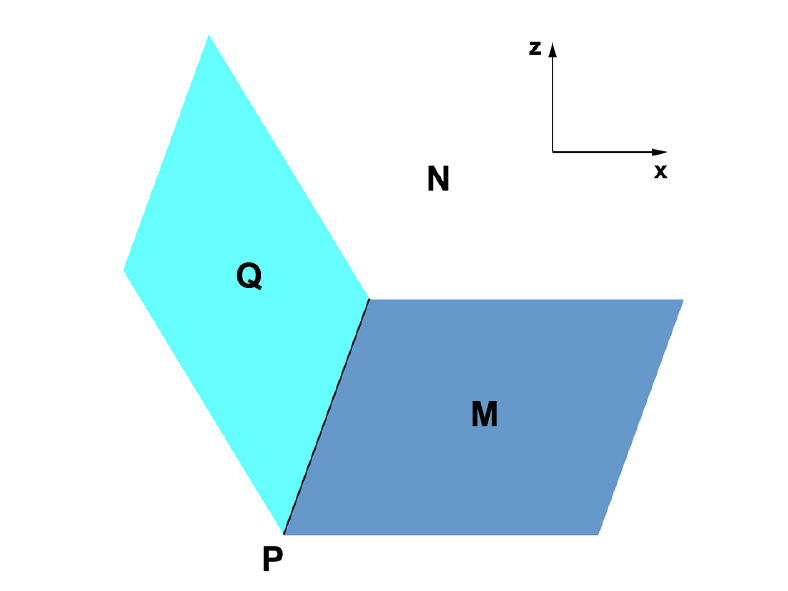}
\caption{Geometry of AdS/BCFT}
\label{MNPQ}
\end{figure}

In this section, we study the AdS/BCFT with CBC. Let us start with
the geometry as shown in Fig.\ref{MNPQ}.  Takayanagi \cite{Takayanagi:2011zk}
proposes to extend the $d$ dimensional manifold $M$ to a $d+1$
dimensional asymptotically AdS space $N$ so that $\partial N= M\cup
Q$, where $Q$ is a $d$ dimensional manifold which satisfies $\partial
Q=\partial M=P$. A central issue in the construction of the
AdS/BCFT is the determination of the location of end-of-the-world brane $Q$.
It turns out the location of $Q$ can be fixed by imposing suitable BCs.
\cite{Takayanagi:2011zk} proposes to choose
NBC, which produces many elegant results and has passed several
non-trivial tests \cite{Nozaki:2012qd,Fujita:2011fp}.
Moreover \cite{Miao:2017aba}, it verifies an universal relation between
Casimir effects and Weyl anomaly, which has been shown to hold
exactly in field theory.
In this paper, we consider AdS/BCFT with the CBC (\ref{CBC}).

Before we start, it is instructive to consider
the simplest vacuum solution of AdS/BCFT with the bulk metric given by
a part of $AdS_{d+1}$:
\begin{eqnarray}\label{CBCgeneralmetric}
  ds^2=dr^2+\cosh^2 r \; \frac{dw^2+\sum_{a=1}^{d-1}dy_a^2}{w^2},\ \
  -\rho \le r <\infty,
\end{eqnarray}
and with the end-of-the-world brane $Q$ situated at
\be \label{Qgen}
r=-\rho.
\ee
Here
$\rho$ is a constant determined by
$T=(d-1)\tanh\rho$.
Note that the extrinsic curvature actually satisfies the
constraint \eq{NBC} and the action variation \eq{variation} vanishes
automatically. As a result, the vacuum solution \eq{CBCgeneralmetric},
\eq{Qgen} is actually
a solution to
AdS/BCFT independent of the type of
 boundary conditions imposed. In particular, it is
a solution to the AdS/BCFT with NBC \cite{Takayanagi:2011zk},
DBC  \cite{Miao:2018qkc}
as well as the CBC \eq{CBC}.
For convenience of later use, the solution can also be written in the Poincare
coordinates by performing the coordinate transformations
\begin{eqnarray}\label{coordinatetransformations}
z=w/\cosh r,\ \ x= w \tanh r.
\end{eqnarray}
The metric \eq{CBCgeneralmetric} becomes
\begin{eqnarray}\label{PoincareAdS}
ds^2=\frac{dz^2+dx^2+\sum_{a=1}^{d-1}dy_a^2}{z^2}
\end{eqnarray}
and the embedding function of $Q$ becomes
\begin{eqnarray}\label{AdSQ}
x=-\sinh\rho  \ z.
\end{eqnarray}

Some comments are in order. First, the CBC \eq{CBC}
apply to only the dimensions higher than two, i.e., $d> 2$. That is
because the 1d and 2d space are always conformal flat, so the second
condition
\eq{CBCb} is trivial for $d\le 2$.
Second, in the above we have considered the choice of parameter
$T< (d-1)$ such that the vacuum AdS can be foliated into slices with $Q$
being an AdS space
\eq{CBCgeneralmetric}. It is possible to consider 
other foliations of the AdS space such that
$Q$ is a dS space or a flat space. This corresponds to the
choices of the tension $T= (d-1) \coth \rho > (d-1)$  or $T=(d-1)$ respectively.
In this paper, we focus on the case with AdS fiolation.
The analysis performed in this paper can be easily generalized to these
other cases.


\subsection{Casimir effects, Weyl anomaly and displacement operator}

For our purpose, let us give a quick review of the Casimir effects, Weyl anomaly
and displacement operator for BCFTs.  In the following sections, we will
frequently use the central charges introduced in this subsection.  

It is found in \cite{Deutsch:1978sc} that
the renormalized stress tensor of BCFT is divergent
near the boundary,
\begin{eqnarray}\label{stress0}
\la T_{ij} \ra  =-2 \a_d\frac{ \bar{k}_{ij}}{x^{d-1}}, \quad  x \sim 0,
\end{eqnarray}
where $x$ is the proper distance from the boundary, $\bar{k}_{ij}$ are the
traceless
parts of extrinsic curvatures and
$\alpha$ is a constant which depends only on the kind of BCFT under
consideration.
The coefficient $\a$ fixes the leading 
shape-dependence
of Casimir effects of
BCFTs. 

Remarkably, the authors of \cite{Miao:2017aba} observe that the above Casimir
coefficients are closely related to the central charges of Weyl anomaly. For
example,  there are universal relations 
\begin{eqnarray}\label{relationg}
\a_3=b_2 ,\  \ \a_4=-\frac{b_4}{2},
\end{eqnarray}
where $b_i$ are boundary central charges of Weyl anomaly of 3d BCFT and
4d BCFT
\cite{Jensen:2015swa,Herzog:2015ioa,Fursaev:2015wpa,Solodukhin:2015eca},
respectively
\begin{eqnarray}\label{3dBWA}
 \mathcal{A} =\int_P\sqrt{h}(b_1 \mathcal{R}+b_2 \text{Tr}
  \bar{k}^2), 
\end{eqnarray}
\begin{eqnarray}\label{4dBWA}
  \mathcal{A}=\text{Bulk Weyl anomaly}+\int_{P}\sqrt{h}( b_3 \text{Tr}
  \bar{k}^3 + b_4 C^{ac}_{\ \ \ b c}
  \bar{k}_{\ a}^b).\;\;
\end{eqnarray}
The
intimate connection of the Weyl anomaly to the Casimir effects has
been generalized to higher dimensions \cite{Miao:2018dvm}, to
anomalous currents \cite{Chu:2018ksb,Chu:2018ntx,Chu:2018fpx,Chu:2019rod}
and Fermion condensations \cite{Chu:2020mwx,Chu:2020gwq}.

Due to the boundary, the energy moment tensor of BCFT is no longer
conserved generally. Instead, we have \cite{Billo:2016cpy}
\begin{eqnarray}\label{DT}
\nabla_i T^{ij}=-\delta(x)D^j(y^a),
\end{eqnarray}
where $D^j(y^a)$ is the displacement operator with scaling dimension $\Delta=d$.
The two point function of displacement operator  is given by
 \begin{eqnarray}\label{2pointD}
\la D^x(y)D^x(0) \ra= \frac{C_D}{|y|^{2d}},
\end{eqnarray}
 with $x$ denotes the normal direction,
 $C_D$ is the Zamolodchikov norm, which is a piece of BCFT data
 \cite{Billo:2016cpy}. In particular, we have
 \begin{eqnarray}\label{positiveD}
C_D\ge 0.
\end{eqnarray}
 It is found in \cite{Miao:2018dvm} that there
 is a universal relation
 \begin{eqnarray}\label{universalrelation1}
\alpha_d=\frac{d\Gamma(\frac{d+1}{2})\pi^{\frac{d-1}{2}}}{2(d-1)\Gamma(d+2)}C_D,
\end{eqnarray}
 between the Casimir coefficient (\ref{stress0}) and displacement operator
 (\ref{2pointD}). See also \cite{Herzog:2017kkj,Herzog:2017xha}
 where an equivalent relation is found for $d=3$ and $d=4$.

 \subsection{
   AdS/BCFT with NBC and DBC}

The Casimir coefficient $\a_d$ is determined in terms of
the boundary central charge.
Similar to the case of the bulk
Weyl anomaly \cite{Henningson:1998gx}, one can expect that generally
the boundary Weyl anomaly (and hence the
boundary central charges) can also be  fixed holographically once
the dual gravitational background is specified. This has been shown to be
indeed the case
in \cite{Miao:2017aba,Miao:2018qkc} for 3d and 4d holographic BCFT obeying
NBC and DBC respectively. 
Let us first briefly review this construction.

In \cite{Miao:2017aba,Miao:2018qkc}, the following ansatz of the bulk metric
\begin{eqnarray}\label{bulkmetric}
  ds^2=\frac{1}{z^2}\Big{[} dz^2+dx^2  +\Big(\delta_{ab}-2\epsilon\ x
    \bar{k}_{ab} f(\frac{z}{x})-2\epsilon\ x  \frac{k}{d-1}\delta_{ab}
 \Big)dy^a dy^b+O(\epsilon^2)\Big{]}
\end{eqnarray} 
was considered, where
$k_{ab}$ is a symmetric matrix,  $\bar{k}_{ab}$ its traceless part and
$k$ its trace.
Without loss of generality, one can choose the boundary condition 
\begin{eqnarray}\label{DBConM}
f(0)=1
\end{eqnarray}
at $z=0$ and
the induced metric on $M$ reads \cite{Miao:2017aba}
\begin{eqnarray}\label{metricM}
ds_M^2=dx^2+\big(\delta_{ab}-2\epsilon x k_{ab}\big)dy^a dy^b+O(\epsilon^2).
\end{eqnarray}
Therefore $k_{ab}$ is simply the
extrinsic curvature of the boundary $P$ of the BCFT.
The metric \eq{bulkmetric} is considered perturbatively with
small $k_{ab}$. However the dependence in $z$ is exact at each perturbative
order.
The function $f$ can be 
determined by substituting (\ref{bulkmetric}) into the Einstein equations.
Performing a perturbative expansion in small $k_{ab}$, with  
$\epsilon$ counting the order of perturbations,
we obtain at the order $O(\epsilon)$ a single equation
\begin{eqnarray}\label{EOM}
s \left(s^2+1\right) f''(s)-(d-1) f'(s)=0.
\end{eqnarray}
This can be solved with 
\begin{eqnarray}\label{solutionf}
f(s)=1+\a_d \frac{s^d \,
 {}_2F_1\left(\frac{d-1}{2},\frac{d}{2};\frac{d+2}{2};-s^2\right)}{d},
\end{eqnarray}
where have used the BC (\ref{DBConM}) and $\a_d$ is an integration constant.
From \eq{bulkmetric} and \eq{solutionf}, we obtain the holographic stress tensor
\begin{eqnarray}\label{NBCholoTij}
  T_{ij}
  =d\frac{\d g_{ij}}{z^d}
  = -2 \epsilon \a_{d} \frac{
  \bar{k}_{ij} }{x^{d-1}} +O(\epsilon^2),
\end{eqnarray}
which takes the expected form
\eq{stress0}.
According to section 2.1, the integral constant $\a_d$
is related to the Casimir coefficient \eq{stress0}, the central
charges of Weyl anomaly \eq{relationg} and the norm of displacement
operator \eq{universalrelation1}.  So far $\a_d$ is arbitrary.
This is correct as
we have not specialized to any specific kind of BCFT.

To specify the holographic BCFT, the position of $Q$ needed to be constrained
by imposing a BC on it.
Consider an embedding of $Q$ given by
\begin{eqnarray}\label{Q1}
x=-\sinh\rho \ z + \lambda\epsilon k z^2+ O(\epsilon^2).
\end{eqnarray}
As shown in \cite{Miao:2017aba, Miao:2018qkc}, this is a solution of the NBC
or DBC
if 
$\a_d$ and $\lambda$ are fixed to be:
\begin{eqnarray}
&& \lambda=\frac{\cosh^2\rho}{2 (d-1)}, \label{solutionlambda}\\
  &&\alpha_{Nd}=\frac{-d \cosh ^d\rho}{(-\coth\rho)^d \,
    _2F_1\left(\frac{d-1}{2},
    \frac{d}{2};\frac{d+2}{2};-\text{csch}^2\rho \right)+d
    \cosh ^2\rho \coth\rho},
  \label{aN}\\
  &&\alpha_{Dd}=\frac{-d (-\text{csch}\rho)^{-d}}{\, _2F_1\left(\frac{d-1}{2},
    \frac{d}{2};\frac{d+2}{2};-\text{csch}^2\rho \right)}.\label{aD}
\end{eqnarray}
Note that suitable analytic
continuation of the hypergeometric function should be taken
in order to get smooth function
at $\rho=0$  \cite{Miao:2017aba, Miao:2018qkc}. For example, we have for
$d=3,4$ explicitly, 
\begin{eqnarray}\label{integralconstantg1}
 && 
  \alpha_{N 3}=\frac{2}{\pi+4 \tan ^{-1}\left(\tanh \left(\frac{\rho }{2}
    \right)\right) }, \qquad \alpha_{N 4}=\frac{1}{2(1+\tanh \rho) },\\
&&
  \alpha_{D 3}=\frac{2}{\pi+4 \tan ^{-1}\left(\tanh \left(\frac{\rho }{2}
    \right)\right)+2\text{csch} \rho },\qquad
  \alpha_{D 4}=\frac{\tanh \rho}{(1+\tanh \rho)^2 }. \label{integralconstantg2}
\end{eqnarray}

At this point, one can try to repeat the analysis for the CBC (\ref{CBC}).
However there is a small subtlety.
We note that for $k_{ab}=0$, the unperturbed metric is just the AdS vacuum
and $Q$ is an AdS-slice, which is conformally flat. As such the CBC simply
asserts that the perturbed metric \eq{bulkmetric} must also be conformally flat:
\begin{subequations}\label{CBC34}
  \begin{empheq}[left=\text{CBC}:\empheqlbrace]{align}
&K=\frac{3}{2} T,\quad \qquad C_{ijk} =0, \quad \mbox{for $d=3$}, \label{CBC34a}\\
&K=\frac{d}{d-1}T,\quad C^i{}_{jkl} =0, \quad \mbox{for $d\ge 4$}, \label{CBC34b}
  \end{empheq}
  \end{subequations}
where $C_{ijk}$ and $C_{ijkl}$ are the intrinsic Cotton tensors and Weyl
curvature tensors on $Q$, respectively.  We have
\begin{eqnarray}\label{Cottontensor}
  C_{ijk}=2\nabla_{[i} S_{j]k}, \quad
  C^{ij}_{\ \ kl}=R^{ij}_{\ \ kl}-4 \delta^{[i}_{[k} S^{j]}_{l]},
\end{eqnarray}
where $S_{ij}$ is the Schouten tensor
\begin{eqnarray}\label{Schouten tensor}
S_{ij}=\frac{1}{d-2} \Big(R_{ij}-\frac{1}{2(d-1)}R h_{ij}\Big).
\end{eqnarray}
Note that $C_{ijk}$ and $C^i_{\ jkl}$ are Weyl invariant for $d=3$ and $d\ge 4$,
respectively. 
It is easy to see that
the CBC (\ref{CBC34}) fixes $\lambda$ to be given by the same
(\ref{solutionlambda}). This is easy to understand since 
$\lambda$ is actually fixed by the asymptotic symmetry of AdS, which is
universal and is independent of the BCs \cite{Miao:2017gyt,Chu:2017aab}.
However \eq{CBC34} does not fix the integral constant $\a_d$.
The reason is because, as a result of
\eq{bulkmetric}, \eq{Q1}, \eq{solutionlambda}, the
induced metric on $Q$ is 
\begin{eqnarray}\label{inducedmetriconQ}
 ds_Q^2=\frac{\sinh^2\rho}{x^2}\Big{[} \coth ^2\rho \; dx^2 
   +\big(\delta_{ab}-2\epsilon\ x \bar{k}_{ab} f(-\text{csch}\rho )-
   2\epsilon\ x
   \frac{k}{d-1}\delta_{ab}
 \big)dy^a dy^b+O(\epsilon^2)\Big{]},\;
\end{eqnarray}
and this is conformal equivalent to 
\begin{eqnarray}\label{inducedmetriconQ1}
 ds_Q^2=dx^2+(\delta_{ab}-2\e \ x K_{ab} ) dy^a dy^b+O(\e^2),
\end{eqnarray}
where $K_{ab}$ are some constant tensors related to $k_{ab}$ and
$f(-\text{csch}(\rho ))$, whose exact expressions are not important.
Now as the Cotton tensors and the Weyl  tensors (\ref{Cottontensor})
contain three and two derivatives, respectively. As a result, for
constant $K_{ab}$, we have $C_{ijk}\sim \e^3 K^3$ and $C_{ijkl}\sim
\e^2 K^2$, which vanish at the linear order of $O(\e)$. This means
that the induced metric (\ref{inducedmetriconQ}) on $Q$ is always
conformal flat at the linear order $O(\e)$, and the CBC \eq{CBC34}
does not impose any constraint on the function $f(s)$.
Therefore, unlike the NBC and DBC,
the imposition of the CBC \eq{CBC34}  does not fix the integral
constant $\a_d $ of the bulk solution \eq{bulkmetric}. This does not mean that
there is any problem with the CBC. It just mean that the ansatz \eq{bulkmetric}
for the bulk metric has to be more complicated in the case of CBC.

\subsection{
  AdS/BCFT with CBC}
  
From
the discussions above, it is clear that it is
the constant extrinsic curvature $k_{ab}$  that makes the
induced metric on $Q$ to be trivially conformal flat at the first order of
$\e$.
The condition
\eq{CBC34} will become nontrivial if we consider
more general extrinsic curvature $k_{ab}(y^a)$ which depends on the
coordinates on the brane $Q$. In this case, the Cotton tensors and the Weyl
tensors is generically non-zero at the linear order of $O(\e)$ and the
nontrivial constraint provided by \eq{CBC34} on the bulk metric can fix
the integration constant $\a_d$. 

For simplicity,  we focus on AdS${}_5$/BCFT${}_4$ with the following bulk metric
\begin{eqnarray}\label{bulkmetricResoI}
 ds^2=\frac{1}{z^2}\Big{[} dz^2+dx^2 
+\delta_{ab}dy^a dy^b-4\e x k_{12} (y_3) f(\frac{z}{x}) dy_1 dy_2 +O(\e^2)\Big{]},
\end{eqnarray}
and the embedding function of $Q$
\begin{eqnarray}\label{Q1ResoI}
x=-\sinh\rho \ z + O(\epsilon^2).
\end{eqnarray}
Here $k_{12} (y_3) $ is given by
\begin{eqnarray}\label{k12y3}
k_{12} (y_3) =k^{(0)}_{12} + \k \ y_3 k^{(1)}_{12} +O(\k^2),
\end{eqnarray}
where $\k$ is a small perturbation parameter, and $k^{(0)}_{12}, k^{(1)}_{12} $
are constants.  Note that the above metric is designed so that the intrinsic
Weyl tensors $C_{ijkl}\sim k^{(1)}_{12} \e \k +O(\e^2, \k^2)$ are non-zero on
the brane $Q$. 
Substituting (\ref{bulkmetricResoI}) into the Einstein functions, we get the EOM
of $f(s)$ as (\ref{EOM}), which can be solved as (\ref{solutionf}).
For $d=4$, we have
\begin{eqnarray}\label{ResoIf4d}
f(s)=1-2 \alpha _4+\frac{\alpha _4 \left(s^2+2\right)}{\sqrt{1+s^2}}.
\end{eqnarray}
From the CBC (\ref{CBC34b}), we get one independent equation
at the linear order of $O(\k)$
\begin{eqnarray}\label{ResoICBCf}
 f(-\text{csch}\rho)\  k^{(1)}_{12} \e \k +O(\e^2,\k^2)=0,
\end{eqnarray}
which imposes a non-trivial constraint on $f(s)$. From (\ref{ResoIf4d})
and (\ref{ResoICBCf}), the integral constant is fixed to be
\begin{eqnarray}\label{ResoICBCcentralcharge}
\alpha_{C 4}=\frac{\tanh \rho}{(1+\tanh \rho)^2 },
\end{eqnarray}
where `C' denotes CBC.
Interestingly, $\alpha_{C 4}$ is exactly the same as that of DBC
(\ref{integralconstantg2}). From
\eq{bulkmetricResoI}, \eq{ResoIf4d} and \eq{NBCholoTij},
we arrive at the holographic stress tensor
\begin{eqnarray}\label{ResoICBCTij}
T_{12}=-2\e\ \alpha_{C 4} \frac{k_{12} (y_3)}{x^3}+ O(\e^2,\k^2).
\end{eqnarray}
As we reviewed in section 2.1,
$\alpha_{C 4}$ gives the central charge $b_4$ of
the boundary Weyl anomaly.

We remark that the ansatz \eq{bulkmetricResoI},
\eq{Q1ResoI} can be naturally generalized to higher dimensions. In fact,
a more general ansatz would be
\begin{eqnarray}\label{bulkmetricResoIgeneral}
 &&ds^2=\frac{1}{z^2}\Big{[} dz^2+dx^2 
    +\big(\delta_{ab}-2\epsilon\ x   \bar{k}^{(0)}_{ab} f(\frac{z}{x}) \big)
    dy^a dy^b-2\e \k \ x y_a H_{abc}(\frac{z}{x}) dy^b dy^c
+O(\epsilon^2, \k^2)\Big{]},\nonumber\\
 &&\\
  && Q: x=-\sinh\rho \ z + \e \k \ \lambda_a z^2 y^a+ O(\epsilon^2,\k^2),
  \label{QResoIgeneral}
\end{eqnarray}
where $x^i=(x,y^a)$, $\lambda_a$ and $H_{abc}(s)$ are constants and
functions to be determined. For simplicity,  we have set the trace
$k=0$ above. 
However, the price to pay for the
ansatz \eq{bulkmetricResoIgeneral}
(and also for \eq{bulkmetricResoI})
is that they depends on too many coordinates
$(z,x,y^a)$, and this makes the analysis of the Einstein equation
and the AdS/BCFT very complicated.  

In this regard, it is instructive to consider the bulk solution in the
coordinates system $(r,w,y_a)$. For simplicity, let us
focus on the case with vanishing traces of extrinsic curvatures, i.e, $k=0$.
Applying the coordinate transformations (\ref{coordinatetransformations}),
the perturbed metric
(\ref{bulkmetric}) becomes
\begin{eqnarray}\label{bulkmetricrw}
  ds^2= dr^2+\cosh^2(r) \frac{dw^2+\big(\delta_{ab}-2\epsilon W(w)
    F(r) \bar{k}_{ab} 
 \big)dy^a dy^b}{w^2}
+O(\epsilon^2),
\end{eqnarray}
where 
\begin{eqnarray}\label{rwW}
&&W(w)=w,\\
&&F(r)=\tanh(r) f(\text{csch}r),\label{rwF}
\end{eqnarray}
and $f(s)$ is given by (\ref{solutionf}).  The embedding
function of $Q$ (\ref{Q1}) becomes
\begin{eqnarray}\label{Q2}
r=-\rho+O(\epsilon^2).
\end{eqnarray}
Imposing NBC (\ref{NBC}) and DBC (\ref{DBC}), we get respectively
\begin{eqnarray}\label{rwNBC}
&& \text{NBC}: \ F'(-\rho)=0,\\
&& \text{DBC}: \ F(-\rho)=0. \label{rwDBC}
\end{eqnarray}
One the other hand, imposing CBC (\ref{CBC34}) we obtain
\begin{eqnarray}\label{rwCBC}
\text{CBC}: 
\begin{cases}
\  W^{(3)}(w) F(-\rho)=0,&\ d= 3,\\
\  W^{(2)}(w) F(-\rho)=0,&\ d\ge 4,
\end{cases}
\end{eqnarray}
where $(n)$ denotes the
$n$th derivative.
Note that Cotton tensors (\ref{Cottontensor}) contain three
derivatives and Weyl tensors include two derivatives, that is why we
have $W^{(3)}(w) $ for $d=3$, while $W^{(2)}(w) $ for $d\ge 4$ above.
Due to (\ref{rwW}), CBC (\ref{rwCBC}) is automatically satisfied and
do not impose any constraint on $F(-\rho)$. As a result, one cannot
fix the integral constant in $F(\r)$ (\ref{rwF}), or equivalently, in
$f(s)$ (\ref{solutionf}).

To resolve this issue, one may  consider a more general non-constant $k_{ab}$
as before. However, it is clear from (\ref{rwCBC}) that the problem can
also be resolved if $W^{(3)}(w)$ and $W^{(2)}(w)$ are non-zero. 
In fact, the choice
(\ref{rwW}) $W(w)=w$ is quite special and can be made more
general. Substituting (\ref{bulkmetricrw}) together with (\ref{rwF},
\ref{solutionf}) into Einstein equations, we obtain the EOM of $W(w)$:
\begin{eqnarray}\label{EOMW}
w^2 W''(w)-(d-2) w W'(w)+(d-2) W(w)=0,
\end{eqnarray}
which can be solved as
\begin{eqnarray}\label{rwWgeneral}
W(w)=
\begin{cases}
\  w(1+ \k \log w),&\ d= 3,\\
\  w(1+ \k \ w^{d-3}),&\ d\ge 4,
\end{cases}
\end{eqnarray}
where $\k$ is an arbitrary constant. For non-zero $\k$, $W^{(3)}(w)$ and
$W^{(2)}(w)$ are indeed non-vanishing. As a result, from (\ref{rwCBC}) we obtain 
\begin{eqnarray}\label{rwCBCgeneral}
\text{CBC}: F(-\rho)=0,
\end{eqnarray}
which is exactly the same as DBC (\ref{rwDBC}).  This is not
surprising. After all, keeping the induced metrics invariant (DBC) is
a special case which keeps the induced metrics conformal invariant
(CBC). However it should be stressed that, in general, CBC and DBC are
different.  As a result,  the CBC  (\ref{rwCBCgeneral}) fixes the
integral constant to be the same as the one (\ref{aD}) of DBC.

Some comments are in order. {\bf 1}. For general $W(w)$, CBC is
well-defined and yields the same integral constants/central charges
(\ref{aD}) as DBC.  {\bf 2}. Although CBC and DBC have the same
$\a_d$
central charge, as we will show in the section 4,
the locations of end-of-the-world brane $Q$ are different for the two BCs and
they define different holographic BCFT in general.
{\bf 3}.  For the general $W(w)$ of (\ref{rwWgeneral}), the induced metric on
the AdS boundary $M$ ($z=w/\cosh r=0$) becomes
\begin{eqnarray}\label{metricMgeneralb}
  ds_M^2=dx^2+\Big(\delta_{ab}-2\epsilon (x+\k \ x^{d-2}) k_{ab}\Big)dy^a dy^b
  +O(\epsilon^2),
\end{eqnarray}
where $x^{d-2}$ should be understood as $x \log x$ for $d=3$ and we
have used $F(\infty)=f(0)=1$ from (\ref{rwF},\ref{DBConM}).
For $d>3$, it is clear from (\ref{metricMgeneralb}) that $\k$
does not affect the value of extrinsic curvatures
\begin{eqnarray}\label{kabgeneralb4d}
\hat{k}_{ab}=-\frac{1}{2}\partial_x g_{ab}|_{x=0}=k_{ab}.
\end{eqnarray}
While for $d=3$, we have
\begin{eqnarray}\label{kabgeneralb3d}
\hat{k}_{ab}=-\frac{1}{2}\partial_x g_{ab}|_{x=0}=(1+ \k \log x)  k_{ab} |_{x=0},
\end{eqnarray} 
which seems to be ill-defined unless $\k=0$.
Note that we have the freedom to Weyl rescale the metric for a BCFT
\begin{eqnarray}\label{metricMgeneralb3d}
  ds_M^2\sim G^2(x)\Big(dx^2+\Big(\delta_{ab}-2\epsilon (x+\k \ x\log x)
  k_{ab}\Big)dy^a dy^b\Big)+O(\epsilon^2).
\end{eqnarray}
For $G(x)=1/(1+\k \log x)$,
one can indeed get a well-defined extrinsic curvature.
\begin{eqnarray}
  \label{kabgeneralb3dp}
  \hat{k}_{ab}=\lim_{x\to 0}\frac{1}{G(x)}\partial_x \Big[
    G(x)^2(x+\k \ x\log x) \Big]= k_{ab}.
\end{eqnarray} 
In fact, a simpler way to get a well-defined extrinsic curvature is to
take $\k$ as a regularization parameter. We take $\k$ small but finite
at the beginning and set it to zero at the end of
calculations. Another regularization is that we focus on the case
$d>3$ and derive the central charges by performing the analytical
continuation $d\to 3$ at the end.  From (\ref{aD}), it is clear that
the analytical continuation $d\to 3$ is well-defined. {\bf 4}. For
more general background metrics such as black strings, the natural BC
on the horizon yields $\k\ne 0$ and thus the well-defined CBC
(\ref{rwCBCgeneral}) is obtained without any arbitrariness. Please see
the next section for details.

\section{
  AdS/BCFT with CBC in
 more general background}

In the above section, we have focused on the perturbations around a vacuum
AdS background with $Q$ being a AdS space.
In this section, we study holographic BCFT with CBC
in more general backgrounds
with $Q$ given by a deformation of the AdS black hole.
In general, the induced metric on $Q$ is no longer conformally flat
and the CBC (\ref{CBC34})
should be replaced by
\begin{subequations}  \label{CBCgeneralbackground}
  \begin{empheq}[left=\text{CBC}:\empheqlbrace]{align}
    & K=\frac{3}{2} T, \quad \qquad \delta C_{ijk}=0,
    \quad \mbox{for $d=3$}, \label{CBCgeneralbackgrounda}\\
    & K=\frac{d}{d-1} T, \quad   \delta C^i_{\ jkl}=0,
    \quad \mbox{for $d\ge 4$}, \label{CBCgeneralbackgroundb}
  \end{empheq}
\end{subequations}
since a conformal rescaling of the metric does not affect the Cotton tensor and
the Weyl tensor. 
A more covariant expression is that we keep conformally invariant all
of the scalars constructed from the intrinsic Cotton tensors and Weyl
tensors on $Q$. For example, we have $\delta (|g| C_{ijl}C^{ijk})=0$
for $d=3$, $\delta ( |g|^{2/d}C_{ijkl}C^{ijkl})=0$ for $d\ge 4$ and so
on.

Since CBC take different forms for $d=3$ and $d>3$, we discuss them
separately below.  The main purpose of this section is to show that,
in the general background, CBC is well-defined and can determine all
of the integral constants and thus the central charges of the Weyl
anomaly.

\subsection{3d BCFT}

Let us start with the case $d=3$ and consider the following ansatz 
\begin{eqnarray}\label{bulkmetric3dBH}
  &&ds^2= dr^2+\frac{\cosh^2(r) }{w^2} \Big[ \frac{dw^2}{1-{w^2}/{w_h^2}}
  - (1-{w^2}/{w_h^2}) dt^2+ dy^2
  \nonumber\\ &&\ \ \ \ \ \ \ \ \ \ \ \ \ \ \ \ \ \ \ \ \ \ \ \ \ \ \ \ \ \ \
  -4\epsilon k_{ty}(1-{w^2}/{w_h^2})  W(w) F(r)dt dy \Big]+O(\epsilon^2),
  \nn \\
&& Q: \ r=-\rho+O(\epsilon^2), \label{bulkQ3dBH}
\end{eqnarray}
where $k_{ty}$ is the extrinsic curvature and $w_h$ denotes the
location of horizon. Note that the induced background geometry on $Q$
is a BTZ black hole.  Substituting (\ref{bulkmetric3dBH}) into
Einstein equations and separating variables, we get
\begin{eqnarray}\label{EOM3dBHF}
&&\frac{F''(r)+3 \tanh (r) F'(r)}{F(r) \text{sech}^2(r)}=\g,\\
&& \frac{w \left(w \left(w^2-w_h^2\right) W''(w)
    +\left(w_h^2+3 w^2\right) W'(w)\right)}{w_h^2 W(w)}=\g,
  \label{EOM3dBHW}
\end{eqnarray}
where $\g$ is the constant of separation.
In general, $\g$ can have
non-trivial dependence on $w_h$. Physically we require that
the induced metric on the AdS boundary is
given by (\ref{metricM}) in the large $w_h$ limit. 
As the LHS of the EOM of $F(r)$ is independent of $w_h$,  the simplest solution
is obtained if $\g$ is independent of $w_h$, i.e.  
\begin{eqnarray}\label{lambda3dBH} 
\g=1.
\end{eqnarray}
More general solutions are possible as long as $\g =1$ in the limit of large
$w_h$. We will focus on the choice \eq{lambda3dBH} here, where in this case
$F$ is given by \eq{rwF}
as in section 2.3 before and the equation (\ref{EOM3dBHW})
has the solution
\begin{eqnarray}\label{W3dBH} 
W(w)=c_1 G_{2,2}^{2,0}\left(\frac{w^2}{w_h^2}|
\begin{array}{c}
 0,1 \\
 \frac{1}{2},\frac{1}{2} \\
\end{array}
\right)+c_2 \frac{ w E\left(\frac{w^2}{w_h^2}\right)}{w^2-w_h^2},
\end{eqnarray}
where $G_{2,2}^{2,0}$ is Meijer $G$ function and $E$ denotes the
complete elliptic integral. Recall that there is a BTZ black hole on
the brane $Q$. We impose the natural BC on the black hole horizon
\begin{eqnarray}\label{naturalBC} 
W(w_h) \ \text{is finite},
\end{eqnarray}
which yields
\begin{eqnarray}
c_2=0.
\end{eqnarray}
Expanding (\ref{W3dBH}) around $w=0$, we have 
\begin{eqnarray}\label{c23dBH}
W(w)= c_1 w \frac{1+\log w-\log 4 w_h }{\pi  w_h}+O(w^2).
\end{eqnarray}
Comparing with (\ref{rwWgeneral}), we get
\begin{eqnarray}\label{c1b3dBH} 
c_1=\frac{\pi w_h}{1-\log 4w_h}, \quad \k=\frac{1}{1-\log 4w_h}.
\end{eqnarray}
Thus the natural BC on the horizon yields a non-zero $\k$. As we have
discussed in section
2.3, the non-zero $\k$ yields a well-defined CBC.  

Now imposing the CBC (\ref{CBCgeneralbackground}) on $Q: r=-\rho$, we get
two independent equations:
\begin{eqnarray}\label{3dBHFCBCeq1} 
&&  \frac{w \left(w^2-w_h^2\right) W''(w)+\left(w_h^2+3 w^2\right)
    W'(w)}{w_h^4} F(-\rho)=0, \\ \label{3dBHFCBCeq2} 
&&   \frac{\left(9 w^2-5 w_h^2\right) W'(w)+\left(w^2-w_h^2\right)
    \left(\left(w^2-w_h^2\right) W^{(3)}(w)+7 w W''(w)\right)}{w_h^4} F(-\rho)=0, 
\end{eqnarray}
which reduce to (\ref{rwCBC}) in the large $w_h$ limit.  Substituting
$W(w)$ (\ref{W3dBH}) with $c_2=0$ into the above equations, we find
that the coefficients of $F(-\rho)$ are non-zero.
As a result, 
\eq{3dBHFCBCeq1} and \eq{3dBHFCBCeq2} are satisfied if $F(-\rho)=0$. This
is precisely the same as (\ref{rwCBCgeneral}) before and give rises to a 
well-defined CBC which fixes all of the
integral constants and  the central charges of the Weyl anomaly.

Some comments are in order.
{\bf 1}. EOM, BCs and solutions of $F(r)$ are the same for
the
AdS background (\ref{bulkmetricrw}) and the more
general background (\ref{bulkmetric3dBH}).  As a result, the integral
constants (central charges) $\a_d$  are
given by
the same (\ref{aD}) for these
two backgrounds. This is consistent with the fact that holographic
boundary
central charges are independent of the bulk solutions.
{\bf 2}.  Recall
that from (\ref{c1b3dBH}) we have $\k\to 0$ in the limit $w_h\to
\infty$. The metric (\ref{bulkmetric3dBH}) considered in this section
provides an exact realization of the regularization discussed around
(\ref{kabgeneralb3d}) of section
2.3. We
can first consider the metric
(\ref{bulkmetric3dBH}) with a large but finite $w_h$, which yield a
well-defined CBC (\ref{rwCBCgeneral}) and fix the integral
constants. Then we set $w_h\to \infty$ and thus $\k \to 0$, which give a
well-defined extrinsic curvature (\ref{kabgeneralb3d}).
{\bf  3}. Finally, we remark that for the
general metric (\ref{bulkmetric3dBH}) with finite $w_h$,
the induced spacetime on the AdS boundary $M$
is not flat.

\subsection{4d BCFT}

Let us go on to discuss the case $d\ge 4$. For simplicity, we focus on
$d=4$ in this subsection and the generalization to higher
dimensions is straightforward. We take the following ansatz 
\begin{eqnarray}\label{bulkmetric4dBH}
  &&ds^2= dr^2+\frac{\cosh^2(r) }{w^2} \Big[ \frac{dw^2}{1-{w^3}/{w_h^3}} -
  (1-{w^3}/{w_h^3}) dt^2+ dy_1^2+dy_2^2 \nonumber\\
  &&\ \ \ \ \ \ \ \ \ \ \ \ \ \ \ \ \ \ \ \ \ \ \ \ \ \ \ \ \ \ \
  -4\epsilon k_{12}W(w) F(r)dy_1 dy_2 \Big]+O(\epsilon^2)\\
&& Q: \ r=-\rho+O(\epsilon^2), \label{bulkQ4dBH}
\end{eqnarray}
where $k_{12}$ denotes the extrinsic curvature and the first line of
(\ref{bulkmetric4dBH}) is the metric of a black string. Substituting
(\ref{bulkmetric4dBH}) into the Einstein equations and separate the
variables as before, we obtain
\begin{eqnarray}\label{EOM4dBHF}
&&\frac{F''(r)+4 \tanh (r) F'(r)}{F(r) \text{sech}^2(r)}=\g,\\
  &&\frac{w \left(w \left(w^3-w_h^3\right) W''(w)+\left(2w_h^3+w^3\right)
    W'(w)\right)}{w_h^3 W(w)}=\g, \label{EOM4dBHW}
\end{eqnarray}
where $\g$ is a constant. Similar to the case of 3d BCFT,
a simple solution where 
$F(r)$ is independent of $w_h$ is obtained if
\begin{eqnarray}\label{lambda4dBH}
\g=2.
\end{eqnarray}
In this case $F(r)$ is given by \eq{rwF}, \eq{solutionf} as in section 2.

Solving (\ref{EOM4dBHW}) with $\g=2$, we get
\begin{eqnarray}\label{W4dBH} 
  W(w)=c_1 w \, _2F_1\left(\frac{1}{3},\frac{1}{3};\frac{2}{3};
  \frac{w^3}{w_h^3}\right)+c_2 w^2 \, _2F_1\left(\frac{2}{3},\frac{2}{3};
  \frac{4}{3};\frac{w^3}{w_h^3}\right),
\end{eqnarray}
Imposing the natural BC (\ref{naturalBC}) on the black hole horizon $w=w_h$,
we obtain
\begin{eqnarray}\label{c24dBH} 
  c_2=-\frac{\Gamma \left(\frac{2}{3}\right)^3}{\Gamma
    \left(\frac{1}{3}\right)^2
    \Gamma \left(\frac{4}{3}\right) w_h} c_1.
\end{eqnarray}
Expanding (\ref{W4dBH}) with (\ref{c24dBH}) around $w=0$, we have
\begin{eqnarray}\label{WonM4dBH} 
  W(w)=c_1 w-\frac{c_1 w^2 \Gamma \left(\frac{2}{3}\right)^3}{w_h
    \Gamma \left(\frac{1}{3}\right)^2 \Gamma \left(\frac{4}{3}\right)}
  +O\left(w^3\right)
\end{eqnarray}
and hence
\begin{eqnarray}\label{c1b4dBH} 
  c_1=1, \quad  \k=\frac{-\Gamma \left(\frac{2}{3}\right)^3}
  {w_h \Gamma \left(\frac{1}{3}\right)^2 \Gamma \left(\frac{4}{3}\right)}
\end{eqnarray}
when \eq{WonM4dBH} is matched with (\ref{rwWgeneral}).
Similar to the 3d case, the natural BC on the horizon yields a
non-zero $\k$, and $\k$ vanishes in the large $w_h$ limit. Note that the
BC (\ref{rwWgeneral}) is chosen for $W(w)$ so that the induced metric
on the AdS boundary is given by (\ref{metricMgeneralb}), which makes
$k_{12}$ of (\ref{bulkmetric4dBH}) to be really an extrinsic
curvature.

Now imposing the CBC (\ref{CBCgeneralbackground}) on
the end-of-the-world brane $Q$, we get two independent equations
\begin{eqnarray}\label{4dBHFCBCeq1} 
&& \frac{w  W(w)}{w_h^3} F(-\rho )=0,\\ \label{4dBHFCBCeq2} 
&&  W''(w) F(-\rho )=0,
\end{eqnarray}
which agree with (\ref{rwCBC}) in the large $w_h$ limit.  From
(\ref{W4dBH}), (\ref{c24dBH}), it is clear that $W(w)$ and $W''(w)$ are
non-zero. As a result, (\ref{4dBHFCBCeq1}) yields $F(-\rho)=0$, which
is
the same (\ref{rwCBCgeneral}) as before. It gives a well defined CBC
and  fixes all the integral
constants (central charges) in $F(r)$.

To summarize, we have shown in this section that the more general
metrics (\ref{bulkmetric3dBH}), (\ref{bulkmetric4dBH}) yield a
well-defined CBC without any arbitrariness. In the large horizon limit
$w_h\to \infty$, we have $\k \to 0$.
In this sense we can think of the metrics  (\ref{bulkmetric3dBH}),
(\ref{bulkmetric4dBH}) as ones providing an
explicit
regularization discussed in section 2.3: that we first take $w_h$
large but finite to get a well-defined CBC, and then send $w_h\to
\infty$ to have a well-defined extrinsic curvature. Since the central
charges are independent of $w_h$ and $\k$, the regularization of
section 2.3 is well-defined.

\section{AdS/BCFT with CBC up to the
  2nd order of perturbations}

In the previous sections, we have focused on the linear perturbations in
the extrinsic curvature around
an AdS background and a black string background. At this order, the coefficient
$\a_d$ (and hence the bulk background) as well as the location of $Q$ 
are found to be the same for CBC and DBC. To detect and determine the other
central charges of the theory,  analysis beyond the linear perturbation
is needed. 

In this section, we will analysis the AdS/BCFT up to
the second order of perturbations. Since all of the central charges of 3d
BCFTs have already been obtained
through the inclusion of linear perturbations
\footnote{For 3d BCFTs, Weyl anomaly
  (\ref{3dBWA}) contains two boundary central charges. The A-type
  central charge $b_1$ is given by $b_1=\sinh(\rho)$
  \cite{Nozaki:2012qd}, which is derived by a pure AdS. And the B-type
  central charge $b_2$ is given by $b_2=\alpha_3$
  (\ref{integralconstantg1}), (\ref{integralconstantg2}),
  which has been  derived by considering
  the linear perturbations around an AdS.}, we will focus on 4d
BCFTs in this section.  We will derive the B-type boundary central charges
$b_3$ and $b_4$, and find that they are the same
for CBC and DBC.  Although the
central charges are the same, the embedding functions of $Q$ are
different for CBC and DBC. The calculations of this section are quite
complicated. Readers who are not interested
in the details can skip
this section.  The main results of this section are summarized in
table 1 and table 2.

At the second order of perturbations,
we find that it is more convenient to work
with the Poincare coordinates $(z,x,y_a)$ as it is
easier to solve Einstein equations in these coordinates. Consider the
following  ansatz for the metric
\begin{eqnarray}\label{bulkmetricwithb}
  && ds^2=\frac{1}{z^2}\Big{[} dz^2+\left(1+\e^2 x^2\left(X(\frac{z}{x})+
    \k x \hat{X}(\frac{z}{x}) \right) \right)dx^2 \nonumber\\
    &&+\left(\delta_{ab}-2\epsilon\ x\left(1+ \k\sqrt{x^2+z^2} \right)
    \bar{k}_{ab} f(\frac{z}{x}) +\e^2 x^2 \left( Q_{ab}(\frac{z}{x})+
    \k x \hat{Q}_{ ab}(\frac{z}{x}) \right)
 \right)dy^a dy^b\Big{]} \nn\\
 &&+O(\e^3, \k^2), 
\end{eqnarray}
and
for the embedding function of $Q$
\begin{eqnarray}\label{Q1withb}
x=-\sinh\rho \ z + \e^2 z^3 (d_1+ \k z\ d_2)+O(\e^3, \k^2).
\end{eqnarray}
Here  $d_1$, $d_2$ are constants and 
$X(s)$, $\hat{X}(s)$, $f(s)$, $Q_{ab}(s)$, $\hat{Q}_{ab}(s)$ are functions
to be determined,
$\e$ is a small perturbation parameters denoting the order of
the small  extrinsic curvature $k_{ab}$, and we have
assumed,
for simplicity, that the trace of extrinsic curvature vanishes $k=0$.
Similar to what we have seen in section 2.3, we have introduced a
small parameter $\k$
whose purpose is to complicate the ansatz \eq{bulkmetricwithb} sufficiently
enough to make the CBC nontrivial so that it can be used to fix all 
of the integral constants in the solution.
$\k$ is free and it can be considered as a regularization parameter since it
can be set to zero at the end of the calculations and \eq{bulkmetricwithb}
\eq{Q1withb} still solve the Einstein equation and the CBC.
We remark that 
(\ref{bulkmetricwithb}) reduces to the ansatz of \cite{Miao:2017aba}
when $\k=0$.
The arrival at this form of the ansatz
\eq{bulkmetricwithb}, \eq{Q1withb} with a
non-zero $\k$ has been a nontrivial task
for this part of our work.

We choose the following BCs on the AdS boundary $M$ $(z=0)$
\begin{eqnarray}\label{BConMXQ} 
f(0)=1,\quad X(0)=\hat{X}(0)=\hat{Q}_{ab}(0)=0,\quad Q_{ab}(0)=q_{ab},
\end{eqnarray}
so that the induced metric on $M$ becomes
\begin{eqnarray}\label{metricoMKKorder} 
  ds_M^2=dx^2+\Big(\delta_{ab}-2\e x(1+\k x) \bar{k}_{ab}+\e^2x^2 q_{ab}+ \cdots
  \Big)
  dy^a dy^b,
\end{eqnarray}
which reduces to the standard form of \cite{Miao:2017aba} when $\k=0$.
For simplicity, we further set 
\begin{eqnarray}\label{kabqab} 
  &&\bar{k}_{ab}=\text{diag}( k_1,-k_1,0 ),\ \ \ \ \ \ \ \ \ \ \ \ \
  q_{ab}=\text{diag}( 0,0,0 ), \\
  &&Q_{ab}(s)=\text{diag}( Q_1(s),Q_2(s),Q_3(s) ),\quad \hat{Q}_{ab}(s)=
  \text{diag}( \hat{Q}_1(s),\hat{Q}_2(s),\hat{Q}_3(s) )\label{QabhatQab}.
\end{eqnarray}
With these, one can go on to
solve Einstein equations with the DBC (\ref{BConMXQ}) on the
AdS boundary $M$ and the CBC (\ref{CBC}) on the end-of-the world brane $Q$. The
approach is similar to that of \cite{Miao:2017aba}, so we do not
repeat it here. Please see the appendix for the solutions of $f(s),
X(s),\hat{X}(s), Q_{ab}(s), \hat{Q}_{ab}(s)$ and $d_1,
d_2$. Interestingly, we find that $f(s), X(s),Q_{ab}(s)$ are the same
for CBC and DBC, while the parameter $d_1$ of the embedding function  is
different. The difference is given by
\begin{eqnarray}\label{d1CBC} 
d_{1 \ DBC}- d_{1 \ CBC}=\frac{1}{18} k_1^2 e^{-4 \rho } \cosh \rho  \coth \rho ,
\end{eqnarray}
which shows that CBC and DBC are indeed different BCs. We remark  that DBC
(\ref{DBC}) can fix all of the integral constants even for $\k=0$
\cite{Miao:2018qkc},
and so we can simply set
$\k=0$
and do not need to
care about $\hat{X}(s),\hat{Q}_{ab}(s),d_2$ in this case.

Substituting the metric (\ref{bulkmetricwithb}) into the holographic stress
tensor  (\ref{NBCholoTij}) and taking the limit $\k\to 0$, we obtain
\begin{eqnarray}\label{TxxCBC4dBCFT} 
  &&T_{xx}=-\frac{2\epsilon ^2 k_1^2  \tanh \rho}{x^2 (1+\tanh \rho)^2} ,
  \\ \label{T11CBC4dBCFT} 
  &&T_{11}=\frac{-2 \epsilon k_1 \tanh \rho }{x^3 (\tanh \rho +1)^2}
  +\frac{\epsilon ^2k_1^2  \text{sech}^8 \rho (7 \sinh 2 \rho 
    +\cosh 2 \rho ) (\sinh 6 \rho +\cosh 6 \rho )}
  {3 x^2 (\tanh \rho +1)^8},\nonumber\\
&&\\  \label{T22CBC4dBCFT} 
  &&T_{22}=\frac{2 \epsilon k_1  \tanh \rho }{x^3 (\tanh \rho +1)^2}+
  \frac{\epsilon ^2 k_1^2  \text{sech}^8\rho (7 \sinh 2 \rho +\cosh 2
    \rho ) (\sinh 6 \rho +\cosh 6 \rho )}{3 x^2 (\tanh \rho +1)^8},
  \nonumber\\ 
&&\\ \label{T33CBC4dBCFT} 
  &&T_{33}=-\frac{\epsilon ^2 k_1^2 \text{sech}^8\rho (2 \cosh 2 \rho 
    -\sinh 2 \rho ) (\sinh 6 \rho +\cosh 6 \rho )}{3 x^2
    (\tanh \rho +1)^8},
\end{eqnarray}
plus terms of $O\left(\epsilon ^3\right)$.
According to  \cite{Miao:2017aba}, the stress tensors near the boundary
can be completely fixed by Weyl anomaly. Applying (A.24) and (A.26) of
\cite{Miao:2017aba}, we have
\begin{eqnarray}\label{TijfromWeylanomaly} 
  T_{ij}=b_4 \frac{ \bar{k}_{ij} -2x \e k^l_{(i}\bar{k}_{j)l}
  }{x^3}\e+
  b_4\frac{(n_i n_j-\frac{1}{3}
    h_{ij})\text{Tr}\bar{k}^2}{2x^2}\e^2
  +(-3b_3-2b_4)\frac{k^l_{(i}k_{j)l}-\frac{1}{3}
    h_{ij}\text{Tr}k^2}{x^2}\e^2
\end{eqnarray}
where $b_3, b_4$ are central charges of the Weyl anomaly, $n_i=(1,0,0,0)$,
$h_{ij} = \text{diag} (0, 1,1,1)$ is the boundary metric,
$k_{ij}=\bar{k}_{ij}=\text{diag}(0,k_1,-k_1,0)$ and we have used
$k=C_{ixjx}=0$ above.  Comparing (\ref{TijfromWeylanomaly}) with
(\ref{TxxCBC4dBCFT}-\ref{T33CBC4dBCFT}), we finally derive the
boundary central charges for 4d BCFTs \footnote{ Note that the
  boundary central charges (\ref{chargesfor4dBCFT}) are different from
  those of \cite{Miao:2017gyt,Chu:2017aab}. Since the induced metric
  on $Q$ of this paper and that of \cite{Miao:2017gyt,Chu:2017aab} are
  not conformally equivalent, it is natural that different BCs yield
  different central charges.}
\begin{eqnarray}\label{chargesfor4dBCFT} 
  b_3=\frac{1}{6}-\frac{e^{-4 \rho }}{2},\quad  b_4=-\frac{2 \tanh \rho}
  {(1+\tanh \rho )^2}.
\end{eqnarray}

Some comments are in order. {\bf 1}.  To compare
(\ref{TijfromWeylanomaly}) with
(\ref{TxxCBC4dBCFT})-(\ref{T33CBC4dBCFT}), we have two free parameters
$(b_3, b_4)$ but four equations. It is non-trivial to have the
consistent
solution (\ref{chargesfor4dBCFT}). This can be regarded as a check of
our calculations. {\bf 2}.  Since CBC and DBC have the same bulk
solutions, they have the same central charges (\ref{chargesfor4dBCFT})
too. {\bf 3}.  Let us stress again that CBC and DBC have different
embedding functions of $Q$ (\ref{Q1withb}), (\ref{d1CBC}). Thus they are
different BCs.  {\bf 4}. To end this section, we list the central
charges of the Weyl anomaly (\ref{3dBWA}), (\ref{4dBWA}) for various BCs in
the following table \ref{table1} and table \ref{table2}.

\begin{table}[htbp]
	\centering 
	\caption{Boundary Central Charge of 3d BCFT}  
	\begin{tabular}{|c|c|c|c|}  
		\hline  
		& & & \\[-6pt]  		
		charge&NBC&DBC&CBC\\  		\hline
		& & & \\[-6pt]  
		$b_1$ & $\sinh\rho$ & $\sinh\rho$ & $\sinh\rho$\\  \hline
		& & & \\[-6pt]  
		$b_2$ & $ \frac{2}{\pi+4 \tan ^{-1}\left(\tanh
                  \left(\frac{\rho }{2}
                  \right)\right) },$&$\frac{2}{\pi+4 \tan ^{-1}\left(\tanh
                  \left(\frac{\rho }{2}
                  \right)\right)+2\text{csch} \rho }$&$\frac{2}{\pi+4
                  \tan ^{-1}\left(\tanh \left(\frac{\rho }{2}
    \right)\right)+2\text{csch} \rho }$ \\  \hline
	\end{tabular} \label{table1}
\end{table}

\begin{table}[htbp]
	\centering 
	\caption{Boundary Central Charge of 4d BCFT}  
	\begin{tabular}{|c|c|c|c|}  
		\hline  
		& & & \\[-6pt]  		
		charge&NBC&DBC&CBC\\  		\hline
		& & & \\[-6pt]  
		$b_3$ & $\frac{1}{1+\tanh\rho}-\frac{1}{3}$ &
                $\frac{1}{6}-\frac{e^{-4 \rho }}{2}$ & $\frac{1}{6}-
                \frac{e^{-4 \rho }}{2}$\\  \hline
		& & & \\[-6pt]  
		$b_4$ & $\frac{-1}{1+\tanh\rho}$&$
                \frac{-2 \tanh \rho }{(1+\tanh \rho )^2}$
                &$\frac{-2 \tanh \rho}{(1+\tanh \rho )^2}$ \\  \hline
	\end{tabular} \label{table2}
\end{table}

\section{Metric perturbations for different BCs}

According to \cite{Witten:1998qj}, CBC is elliptic and leads to a well-defined
perturbation theory for gravity. In this section, we analysis the
metric perturbation
with CBC, DBC and NBC imposed respectively
on the end-of-the-world brane $Q$. 


\subsection{Metric perturbations}

Consider the following ansatz of the
perturbation metric and the embedding function of $Q$
 \begin{eqnarray}\label{perturbationmetric}
   &&ds^2=dr^2+\cosh^2r \left( \bar{h}^{(0)}_{ij}(y) + \epsilon H(r)
   \bar{h}^{(1)}_{ij}(y)  \right)dy^i dy^j+O(\epsilon^2),\\
&& Q: \ r=-\rho+O(\epsilon^2), \label{perturbationQ}
\end{eqnarray}
 where $\bar{h}^{(0)}_{ij}(y)$ is the
 unperturbed metric and
 $\bar{h}^{(1)}_{ij}(y)$ denotes the perturbation.
 Our results below for the spectrum of gravitons is independent of the choice
 of the unperturbed metric. Interesting examples of the unperturbed metric
 includes, for example,
 the $d$ dimensional AdS space
 in the RS brane world
 scenario or the AdS black hole for applications in islands.
In terms of bulk metric perturbations,  we have
 \begin{eqnarray}\label{bulkmetricperturbations}
\delta g_{r\mu}=0,\quad \delta g_{ij}=\cosh^2 (r)  H(r) \bar{h}^{(1)}_{ij}(y).
\end{eqnarray}
The usual transverse traceless gauge  
 \begin{eqnarray}\label{gijgauge}
\nabla^{\mu} \delta g_{\mu\nu}=0,\quad  g^{\mu\nu}\delta g_{\mu\nu}=0,
\end{eqnarray}
 reads in the linear order
 \begin{eqnarray}\label{hij1gauge}
D^i \bar{h}^{(1)}_{ij}=0,\quad  \bar{h}^{(0)ij}\bar{h}^{(1)}_{ij}=0,
\end{eqnarray}
where $\nabla^{\mu}$ and $D^i$ are the covariant derivatives with
respect to $g_{\mu\nu}$ and $\bar{h}^{(0)}_{ij}$
respectively.
 In the gauge (\ref{gijgauge}), the linearlized Einstein equations become
 \begin{eqnarray}\label{EOMgABgauge}
 \left(\nabla_{\alpha} \nabla^{\alpha}+2\right) \delta g_{\mu\nu}=0.
\end{eqnarray}
 Substituting (\ref{bulkmetricperturbations}), (\ref{hij1gauge}),
 into (\ref{EOMgABgauge}) and separating variables, we obtain
 \begin{eqnarray}\label{EOMMBCmassivehij}
&& \left(D_i D^i+2-m^2\right)\bar{h}^{(1)}_{ij}(y)=0,\\
   && \cosh^2(r) H''(r)+d \sinh (r) \cosh (r) H'(r) + m^2 H(r)=0,
   \label{EOMMBCmassiveH}
\end{eqnarray}
 where $m^2$
 is the constant of separation.
 The spectrum of fluctuations is determined once the boundary conditions of the
AdS/BCFT are specified.
We impose the standard DBC on the AdS boundary $M$
 \begin{eqnarray}\label{DBConMforH}
H(\infty)=0. 
 \end{eqnarray}
In addition, we impose a BC on $Q$. For the choice of CBC,
the second condition of \eq{CBC34}
or the second condition \eq{CBCgeneralbackground} both
yields
 \begin{eqnarray}\label{HCBCagian}
\text{CBC/DBC}: \quad H(-\rho)=0.
 \end{eqnarray}
 Note that DBC also yields  the same condition \eq{HCBCagian}. This implies
 that the  spectrum of fluctuations are the same for CBC and DBC.
 As for NBC, we have the condition
 \be \label{HNBCagian}
\text{NBC}: \quad  H'(-\rho)=0.
\ee

Each mode of fluctuation of $\bar{h}^{(1)}_{ij}$ represents a massive
graviton of mass $m$
in the theory. It also has the
interpretation as fluctuation mode
\be 
\d K_{ij} =
-\frac{\epsilon}{2}\cosh ^2(\rho ) H'(-\rho ) \bar{h}^{(1)}_{ij} \label{KijonQ}
\ee
of the extrinsic curvature
$K_{ij} = -\frac{1}{2} \partial_r g_{ij}\big|_{r=-\rho}$ of $Q$.
Note that $\d K_{ij}$ vanishes for NBC but is non-trivial for CBC and DBC.
Note also that the perturbation \eq{KijonQ}, while nontrivial, actually
keeps $K$ invariant and this is 
consistent with the CBC and DBC. 
\subsection{Mass spectrum}
\begin{figure}[ht!]
\centering
\includegraphics[width=10cm]{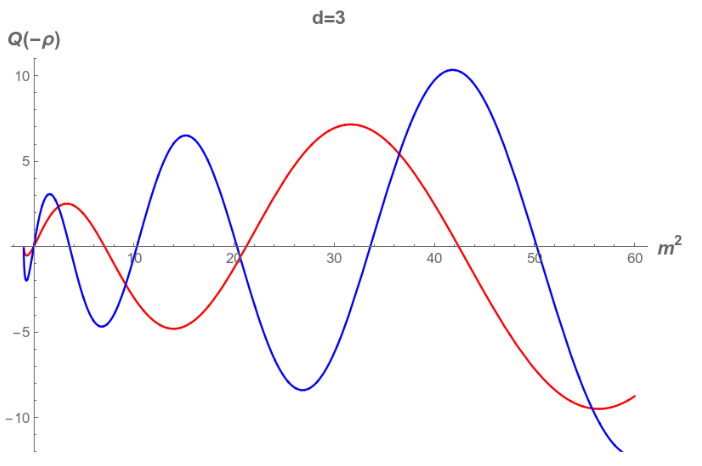}
\includegraphics[width=10cm]{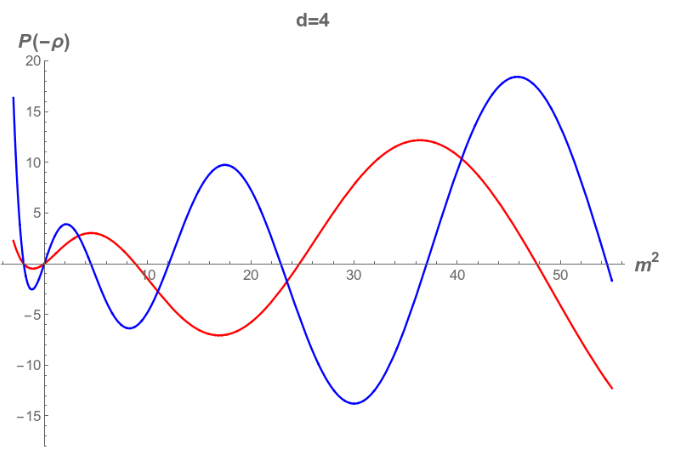}
\caption{Mass spectrum for CBC/DBC with $d=3, 4.$
 Here the roots of $H(-\rho)=0$ gives
  the allowed masses $m^2$ for CBC and DBC. Red (Blue) curve is for
  $\rho =0.1 (1)$.}
\label{mass3d4d}
\end{figure}
Let us
now work out the mass spectrum of the gravitons.
 For generic $m^2$, the general solution to (\ref{EOMMBCmassiveH}) is given
in terms of the two  independent Legendre functions
 $P_{\l}^{\frac{d}{2}}(x)$ and $ Q_{\l}^{\frac{d}{2}}(x)$ by
 \begin{eqnarray}\label{massiveHsolution}
   H(r)=\text{sech}^{\frac{d}{2}}(r) \left(c_1 P_{\l}^{\frac{d}{2}}(\tanh r)
   +c_2 Q_{\l}^{\frac{d}{2}}(\tanh r)\right).
\end{eqnarray}
 Here
 $c_1$ and $c_2$ are integral constants and $\l$ is 
 \begin{eqnarray}\label{aibia1}
\l :=\frac{1}{2} \left(\sqrt{(d-1)^2+4  m^2}-1\right).
 \end{eqnarray}
We note however that for
the special value of $m^2$
\begin{equation}\label{neg-m2}
  m^2 =\begin{cases}
  -\frac{1}{4}((d-1)^2-1),&\ \text{even $d$},\\
  -\frac{1}{4}(d-1)^2,&\ \text{odd $d$},
\end{cases}
\end{equation}
which corresponds to $\l=0$ ($d$ even) and $\l =-1/2$ ($d$ odd) respectively,
\eq{massiveHsolution} no longer gives the general solution since
the Legendre functions $P^n_0(x), Q^{n+\frac{1}{2}}_{-\frac{1}{2}}(x)$ vanish
identically for integer $n$. In this case,
the general solution is given, 
for even $d$,
by
\be \label{H-speca}
H(r) = \text{sech}^{\frac{d}{2}}(r) \left(c_1 e^{\frac{dr}{2}} + c_2 e^{-\frac{dr}{2}}
\right),
\ee
and for odd $d$, by
\be \label{H-specb}
H(r) =  \text{sech}^{\frac{d}{2}}(r) \left(
c_1 P^{\frac{d}{2}}_{-\frac{1}{2}}(\tanh r) + c_2 P^{-\frac{d}{2}}_{-\frac{1}{2}}(\tanh r)
\right).
\ee
As $P^{\pm \frac{d}{2}}_{-\frac{1}{2}}(\tanh r) \sim e^{\pm \frac{dr}{2}}$ for large
$r \to \infty$, it is clear  that (\ref{H-speca},\ref{H-specb})
  cannot satisfy the boundary conditions
(\ref{DBConMforH}, \ref{HCBCagian}) or (\ref{DBConMforH}, \ref{HNBCagian}) 
and so we can rule out \eq{neg-m2} from the spectrum.
Next us now get back to the discussion for generic $m^2$.
The imposition of \eq{DBConMforH} gives
\begin{equation}\label{Htwocase}
H(r)=\begin{cases}
\  c_1 \ \text{sech}^{\frac{d}{2}}(r)  \ P_{\l}^{\frac{d}{2}}(\tanh r),&\
\text{even $d$} ,\\
\  c_2\  \text{sech}^{\frac{d}{2}}(r) \ Q_{\l}^{\frac{d}{2}}(\tanh r),&\
\text{odd $d$}.
\end{cases}
\end{equation}
We note that for $m^2=0$, it is $\l = d/2 -1$ and $H(r)$ as given by
\eq{Htwocase} is identically zero. So massless graviton is 
excluded for all cases, CBC, DBC and NBC, of boundary condition.

Next, let us first consider the boundary condition CBC/DBC \eq{HCBCagian} on
the end-of-the-world brane $Q$.
We get the constraint on
$m^2$:
\begin{equation}\label{constraintonM}
  0 =\begin{cases}
  P_{\l}^{\frac{d}{2}}\left(\tanh \left(-\rho\right)\right),&\ \text{even $d$},\\
  Q_{\l}^{\frac{d}{2}}\left(\tanh \left(-\rho\right)\right),&\ \text{odd $d$}.
\end{cases}
\end{equation}
For $m^2 < -\frac{(d-1)^2}{4}$ outside the  Breitenlohner-Freedman (BF) bound,
there is no solution to \eq{constraintonM} since in this case
$\l = -1/2 + i \a \; (\a > 0)$ becomes complex 
and the Legendre function $P^{\frac{d}{2}}_\l (\tanh(-\r))$
is real and strictly positive, while $Q^{\frac{d}{2}}_\l (\tanh(-\r))$
is purely imaginary and strictly negative for this range of $m^2$.
For  $m^2 \geq -\frac{(d-1)^2}{4}$
 satisfying the BF bound of AdS${}_d$,
 as massless graviton is excluded,
 all the solutions have positive $m^2$.  
In Fig \ref{mass3d4d}, we plot for $d=3,4$ the value of
$P^{\frac{d}{2}}_\l(-\r)$ or $Q^{\frac{d}{2}}_\l(-\r)$
against $m^2$.  The roots of the curves give the admissible
values of   $m^2$.

To get further
understandings of the spectrum, let us study some special cases.
For large $\rho\to \infty$, (\ref{constraintonM}) can be
approximated
by \cite{book}
\begin{equation}\label{constraintonMlarge}
0=\lim_{\rho\to \infty} H(-\rho)\sim \begin{cases}
 -2^{d/2} \pi^{-1}  \sin(\l \pi) \,&\ \text{even $d$} ,\\
 -2^{(d/2)-1} \Gamma(d/2)\cos(\l \pi),&\ \text{odd $d$},
\end{cases}
\end{equation}
which has the roots
\begin{equation}\label{masslargerho}
m^2 \approx 
k (k+d-1), \, \ \ \ \ \text{for large } \rho,
\end{equation}
where $k\ge 1$ are integers. 
On the other hand, in the limit of small $\rho \to 0$, we have 
\begin{equation}\label{constraintonMsmall}
0=\lim_{\rho\to 0} H(-\rho)\sim \begin{cases}
  f_1(m^2)1/\Gamma \left(\frac{1}{4}
  \left(d+3-\sqrt{(d-1)^2+4 m^2}\right)\right),
  \,&\ \text{even $ d$} ,\\
  f_2(m^2) \sin \left(\frac{1}{4} \pi
  \left(d-1+\sqrt{(d-1)^2+4 m^2}\right)\right),&\ \text{odd $d$},
\end{cases}
\end{equation}
where $f_1>0$ and $f_2>0$ are some unimportant positive functions.
From (\ref{constraintonMsmall}), we derive the roots
\begin{equation}\label{masssmallrho}
m^2 \approx 
2 k (2 k+d-1),
\ \ \ \ \text{for small } \rho,
\end{equation}
where $k\ge 1$ are integers.  Remarkably, the spectrum
(\ref{masssmallrho}) of small $\rho$ is included in the spectrum
(\ref{masslargerho}) of large $\rho$ with even mass level $k$.

\begin{figure}[ht!]
\centering
\includegraphics[width=10cm]{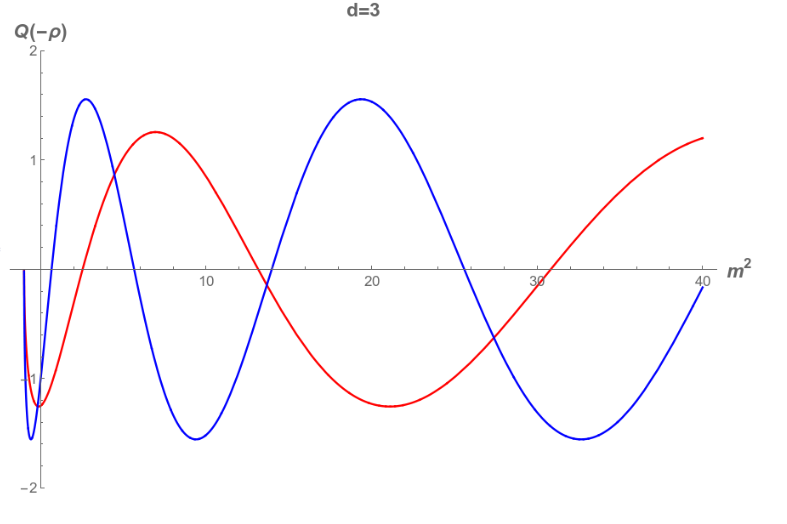}
\includegraphics[width=10cm]{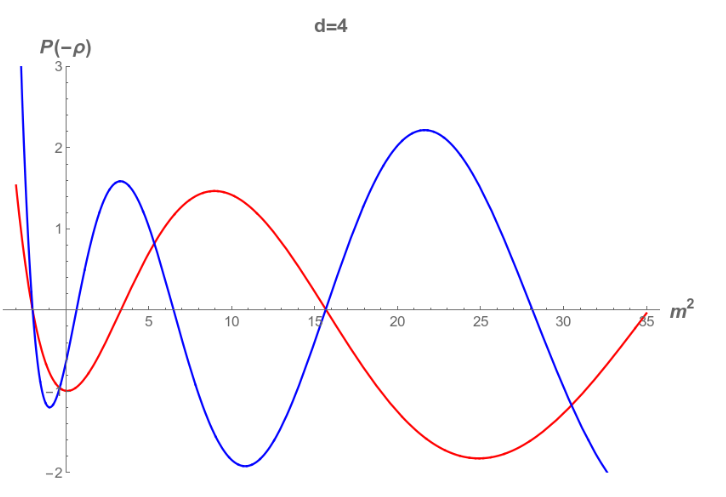}
\caption{Mass spectrum for NBC with $d=3, 4.$
 Here the roots of $H'(-\rho)=0$ gives
  the allowed masses $m^2$ for NBC. Red (Blue) curve is for
  $\rho =0.1 (1)$.}
\label{mass3d4d-NBC}
\end{figure}

Finally, let us work out the metric perturbation spectrum
with
the NBC
\eq{HNBCagian}
imposed on $Q$.
After some simplification, we obtain the following constraints on $m^2$:
\begin{equation}\label{constraintonM-NBC}
  0 =\begin{cases}
  P_{\l}^{\frac{d}{2}-1}\left(\tanh \left(-\rho\right)\right),&\ \text{even $d$},\\
  Q_{\l}^{\frac{d}{2}-1}\left(\tanh \left(-\rho\right)\right),&\ \text{odd $d$}.
\end{cases}
\end{equation}
We remark that while \eq{constraintonM-NBC} for NBC appears to be the
same as \eq{constraintonM-NBC} for CBC/DBC
with $d$ replaced by $(d-2)$, this is not true since
the parameter $\l$ as given by  \eq{aibia1} is not changed.
The analysis of the solutions to \eq{constraintonM-NBC} is similar to
the the discussion above for the case of CBC/DBC.
The perturbation modes are all positive, as seen in the figure
\ref{mass3d4d-NBC}.

Summarizing, the quadratic
metric fluctuations of AdS/BCFT are all massive.
On the end-of-the-world brane, they manifest
as  fluctuation modes \eq{KijonQ} of the extrinsic curvature
$\d K_{ij} $ in the case of CBC and DBC, and as fluctuation modes of the
metric $\d g_{ij}$ in the case of NBC:
\be \label{dynamicalgravity}
\begin{cases} 
  \d K_{ij} = -\frac{\epsilon}{2}\cosh ^2(\rho ) H'(-\rho ) \bar{h}^{(1)}_{ij}
  & \mbox{for CBC and DBC},\\
  \d g_{ij} = \e \cosh^2(\r)  H(-\r)\bar{h}^{(1)}_{ij} & \mbox{for NBC}.
\end{cases}
\ee
The mass spectrum is determined by \eq{constraintonM}
in the case of CBC and DBC,
and \eq{constraintonM-NBC} in the case of NBC.
The spectrum is countable and positive, and we can denote it as $\{ m_n^2 \}$ .

\subsection{Two point functions in BCFT}
As an application for the mass spectrum obtained above, let us consider the
two point functions of operators in BCFT. For simplicity, let us discuss
the case
of primary operator $\cO_\D$ with conformal dimension $\D$. 
We use the Gaussian normal
coordinates $y^i = (w, \vy)$ on $M$ where $w$ is the coordinate in the
normal direction and $\vy = (y^a)$ is the coordinates for the orthogonal slices.
The two-point function in BCFT for primary operators $\cO_i$ with
dimensions $\D_i$ is generally given by 
\be \label{two point primary operators}
\la \cO_1(w,\vy) \cO_2(w',\vy') \ra =
\frac{f(\eta)}{(2w)^{\D_1} (2w')^{\D_2}}, 
\ee
where $f$ is a function of the conformal invariant
\be
\eta = \frac{(\vy-\vy')^2 +(w-w')^2}{w w'}.
\ee
It was shown in \cite{Karch:2018uft}
that 2-point function for the same scalar operators
$(\D_1 = \D_2 =\D$) can be
constructed in terms of the bulk-to-bulk propagators in AdS space
 \footnote{
See also \cite{Bissi:2018mcq,Mazac:2018biw,Reeves:2021sab,Kastikainen:2021ybu} for further work on the correlation functions of BCFT.}. The result is
\be
f(\eta) = (2\D-d)^2 \sum_n C_n^2 \cG_n (\eta),
\ee
where the sum $n$ is over the mass spectrum $\{m^2_n\}$ of the radial equation
\eq{EOMMBCmassiveH}. Here the
conformal block $\cG_n$ is given by the standard bulk-to-bulk AdS${}_d$
propagator
\be
\cG_n (\eta) = \frac{C_{\D_n,d}}{2\D_n-(d-1)}(\eta)^{-\D_n}
{}_2F_1 \left(\D_n,\D_n-\frac{d}{2}+1, 2\D_n-d+2,-\frac{4}{\eta} \right),
\ee
where
$
C_{\D_n,d} =\frac{\G(\D_n)}{\pi^\frac{d-1}{2} \G(\D_n-\frac{d-1}{2})}
$
and $\D_n$ is determined by $\D_n (\D_n -(d-1)) = m_n^2$. 
The numerical factor $C_n$  can be extracted from
the fall off 
\be
H_n (r) = C_n e^{-\D_n r} + O (e^{-(\D_n+2)r})
\ee
at large $r$. With the mass spectrum $\{m_n^2 \}$
and the mode functions $H_n(r)$ we determine
above, the 2-point function for holographic BCFT with CBC, DBC or
NBC imposed is entirely fixed and one can perform a detailed study of
its properties.
Note that the EOM (\ref{EOMMBCmassivehij}) and BCs (\ref{DBConMforH},
\ref{HCBCagian}, \ref{HNBCagian}) of $H_n(r)$ are exactly the same for
the massless scalar and gravity in the bulk. Thus our above results
can help to study the holographic two point function of the scalar
primary operator (\ref{two point primary operators}).  
See appendix for
the discussions of the mass spectrum
for massive bulk scalars. 
 It is also interesting
to extend the analysis to determine the 2-point function for
energy-momentum tensor in AdS/BCFT.
We leave it to future works.

\section{More physical discussions}

In this section, we make more physical discussions
concerning our results.


\subsection{Relation to massive island}


The authors of
\cite{Geng:2020qvw} notice that all reliable calculations of the Page
curve in more than three dimensions have been performed in
gravity theories  with only massive gravitons.
If one attempts to take the limit of zero
graviton mass, any contribution from the islands disappears. In their
following work \cite{Geng:2021hlu}, the authors further argue that
islands might not constitute consistent entanglement wedges in
massless gravity where the Gauss law applies.
On the other hand, the puzzle can be
resolved naturally in massive gravity, where the Gauss law does not
apply. This implies that consistent island formation seems to be 
possible only  in massive gravity.
Their argument is general. So far islands has been studied mainly
with NBC.
Our results show that the gravitons
in the AdS black hole backgrounds with AdS brane are indeed massive for all
three kinds of boundary conditions. 
It will be interesting to study the properties of islands with CBC/DBC as well.


\subsection{
  Consistency of AdS/BCFT boundary conditions}

In this subsection, we argue that CBC/DBC and NBC all define a
consistent theory of AdS/BCFT. Different holographic BCFTs
self-consistently correspond to different boundary conditions. The
reasons are as follows. First, there are
generally more than one
consistent boundary conditions that can be imposed
for a theory. For examples, one can
impose either DBC and NBC for scalar theory,
and either the absolute boundary condition or the relative boundary
condition
for the Maxwell theory. It is natural that there are more than one consistent
boundary conditions for AdS/BCFT
as well.
Second,
we note that the holographic Weyl anomaly obtained with CBC, DBC and NBC
not only admit the correct form in agreement with general classification, but
also the corresponding boundary charges all possess the correct physical
properties.
Take 3d BCFT as an example. The A-type boundary central
charge is listed in Table. 1,
  \begin{eqnarray}\label{A-typecharge}
b_1= \sinh\rho.
 \end{eqnarray}
  From the null energy condition on the brane,
  \cite{Takayanagi:2011zk,Fujita:2011fp} derive
 \begin{eqnarray}\label{rhoUVIR}
\rho_{UV} \ge \rho_{IR},
 \end{eqnarray}
which has a natural geometric interpretation \cite{Miao:2020oey}.
From (\ref{AdSQ}) \cite{Miao:2020oey} notices that the larger $\rho$
is, the closer the brane approaches to the AdS boundary (UV). Thus,
large $\rho$ corresponds to the UV, while
small $\rho$ corresponds
to the IR. From \eq{A-typecharge} and \eq{rhoUVIR},
therefore the
holographic g/c-theorem for the A-type boundary central charge
   \begin{eqnarray}\label{b1UVIR}
b_{1\ UV} \ge b_{1\ IR}
   \end{eqnarray}
is observed.
   This is a strong support for
   the consistency of all 3 kinds of boundary conditions.
   As for the  B-type boundary central charge, they are given by
   (\ref{integralconstantg1},\ref{integralconstantg2}) and are different
   for CBC/DBC and NBC.
   For a unitary
BCFT, the B-type boundary central charge, which is related to the norm
of displacement operator, should be non-negative (see section 2.1 for
more discussions)
   \begin{eqnarray}\label{adallBCs}
\alpha_d\ge 0.
 \end{eqnarray}
   As shown in \cite{Miao:2018qkc},
   this is indeed the
case for all kinds of BCs, as long as the brane tension is
non-negative, i.e., $\rho\ge 0$.
In other word, CBC/DBC and NBC all yield ``correct Weyl
anomaly" with A-type boundary central charges obeying the
holographic g/c-theorem
and non-negative B-type boundary central charges that correspond to
unitary theories.
Third, graviton are massive for all kinds of BCs.
This is another support for various choices of
BCs. For the above reasons, we
believe that the AdS/BCFT with CBC/DBC and NBC are all well-defined.
To get more tests of different BCs, one can study
two-point functions of stress tensor and the bootstrap constraints
\cite{Rychkov:2016iqz,Simmons-Duffin:2016gjk,Poland:2018epd}. These
are non-trivial problems which go beyond the main purpose of this
paper. We leave them to future works.

\section{Conclusions}

In this paper, we have investigated the holographic BCFT with CBC.
The CBC is an interesting BC of gravity, which is elliptic and leads
to a well-defined perturbation theory of `quantum' gravity
\cite{Witten:2018lgb}.
For simplicity, we have focused on the classical gravity.
We derived the massive gravitational modes
for various boundary conditions imposed 
on the end-of-the-world brane $Q$.
Compared
with NBC and DBC, CBC is more subtle. For the 
simplest perturbations with homogeneous extrinsic curvature
around the AdS background, we showed that 
CBC  does not impose any constraint on the
integral constants of the solution and the central charges of
the Weyl anomaly.
Nevertheless the central charges of the Weyl anomaly can be fixed
holographically if one consider more general metric perturbations around
an AdS or a black string background.
In this way, we fix the central charges of Weyl anomaly for
holographic BCFT with
CBC. Interestingly, we find that the central charges are the same for
CBC and DBC, although they are different BCs in general
since they yield
different locations of the end-of-the-world brane.  We also study the
gravitational
perturbations for the different choices of BCs.
Remarkably, we find that there are non-trivial
gravitational dynamics from the extrinsic curvatures, which obeys EOM
of massive gravity at the linear perturbations on a Dirichlet and
conformal brane.
Finally, we analysis the mass spectrum of metric perturbations
and find that there is no
massless mode in the theory
for all kinds of boundary conditions.
Our results show that it should be possible to form island in AdS/BCFT with
CBC/DBC as well.

Many interesting problems are worth exploring.
The existence of massive gravitons
is interesting and it is expected to have interesting 
implications on the BCFT through holography.
In this paper, we have worked out the mass spectrum of the
gravitons. It is  interesting to
study the non-linear dynamics of the metric perturbations in
order to better understand their interactions.
Further study of double holography with CBC, such as wedge
holography is also an interesting problem.
For example, according to the results of this paper, the wedge
holography with CBC yields the correct forms of Weyl anomaly. In
particular, the A-type Weyl anomaly is exactly the same for wedge
holography with
CBC and NBC. Investigation of other aspects of the holography would
be interesting.
For example, how do the perturbation modes affect the
quantum extreme surface and the island formula on the Dirichlet and
conformal branes?
We hope these interesting problems could be
addressed in the near future.

\section*{Acknowledgments}
C. S. Chu acknowledge support of
this work by NCTS and by the grant 110-2112-M-007-015-MY3 of the Ministry of
Science and Technology
of Taiwan.
R. X. Miao thank the support by NSFC grant (No.11905297)
and Guangdong Basic and Applied Basic Research Foundation
(No.2020A1515010900).

\appendix
\section{Second order perturbative solutions to AdS/BCFT with CBC}


In this appendix, we solve the bulk Einstein equations for the ansatz
\eq{bulkmetricwithb} that
include perturbations in the extrinsic curvature up to the second order.
Solving Einstein equations together with the DBC (\ref{BConMXQ}) on
AdS boundary $M$ and the CBC (\ref{CBC34}) on the end-of-brane $Q$, we obtain
\begin{eqnarray}\label{4dBCFTCBCf} 
  &&f(s)=1+2 a_0-\frac{a_0 \left(s^2+2\right)}{\sqrt{s^2+1}},\\
  \label{4dBCFTCBCX} 
  && X(s)=-\frac{k_1^2}{2}  \left(2 a_0 \left(a_0 \left(8 \sqrt{s^2+1}+s^2
  \left(\log \left(s^2+1\right)-4\right)-8\right)-2 s^2+4 \sqrt{s^2+1}
  -4\right)+s^2\right),\nonumber\\
\end{eqnarray}
\begin{eqnarray}\label{4dBCFTCBCQQ1} 
  Q_1(s)&=&a_1 \left(s^2-2 \sqrt{s^2+1}+2\right)-a_0^2 k_1^2 \log
  \left(s^2+1\right)\nonumber\\
  &&+\frac{k_1^2 \left(a_0 \left(-4 s^2+8 \sqrt{s^2+1}-8\right)-s^2
    +\sqrt{s^2+1}-1\right)}{\sqrt{s^2+1}}\nonumber\\
  &&+\frac{a_0^2 k_1^2 \left(-14 s^4+\left(51 \sqrt{s^2+1}-76\right) s^2
    +62 \left(\sqrt{s^2+1}-1\right)\right)}{6 \left(s^2+1\right)^{3/2}},
\end{eqnarray}
\begin{eqnarray}\label{4dBCFTCBCQQ2} 
  Q_2(s)&=&a_2 \left(s^2-2 \sqrt{s^2+1}+2\right)-a_0^2 k_1^2
  \log \left(s^2+1\right)\nonumber\\
  &&+\frac{k_1^2 \left(a_0 \left(-4 s^2+8 \sqrt{s^2+1}-8\right)-s^2
    +\sqrt{s^2+1}-1\right)}{\sqrt{s^2+1}}\nonumber\\
  &&+\frac{a_0^2 k_1^2 \left(-14 s^4+\left(51 \sqrt{s^2+1}-76\right) s^2
    +62 \left(\sqrt{s^2+1}-1\right)\right)}{6 \left(s^2+1\right)^{3/2}},
\end{eqnarray}
\begin{eqnarray}\label{4dBCFTCBCQQ3} 
Q_3(s)&=&\left(a_1+a_2\right) \left(-s^2+2 \sqrt{s^2+1}-2\right)\nonumber\\
&&+\frac{1}{2} k_1^2 \left(a_0 \left(-4 s^2+8 \sqrt{s^2+1}-8\right)-s^2
+4 \sqrt{s^2+1}-4\right)\nonumber\\
&&+\frac{1}{6} a_0^2 k_1^2 \left(15 s^2-20 \sqrt{s^2+1}-\frac{1}{s^2+1}-6
\log \left(s^2+1\right)+21\right),
\end{eqnarray}
\begin{eqnarray}\label{4dBCFTCBChatX} 
  \hat{X}(s)&=&\frac{4}{3} k_1^2 \left(\left(\sqrt{s^2+1}-5\right) s^2
  +\sqrt{s^2+1}-1\right)\nonumber\\
  &&-\frac{2}{3} a_0^2 k_1^2 \left(\left(4 \sqrt{s^2+1}+6\right) s^2
  -8 \sqrt{s^2+1}-6 \left(s^2+1\right) \log \left(s^2+1\right)
  +8\right)\nonumber\\
  &&+\frac{2}{3} a_0 k_1^2 \left(\left(8 \sqrt{s^2+1}-15\right) s^2
  +8 \left(\sqrt{s^2+1}-1\right)+3 \left(s^2+1\right)
  \log \left(s^2+1\right)\right),\;\;\;\;\;
\end{eqnarray}
\begin{eqnarray}\label{4dBCFTCBChatQQ1} 
  \hat{Q}_1(s)&=&\frac{1}{2} \ha_1 \left(\left(2 \sqrt{s^2+1}-3\right) s^2
  +2 \left(\sqrt{s^2+1}-1\right)\right)\nonumber\\
&&+\frac{4}{3} k_1^2 \left(-2 s^2+3 \sqrt{s^2+1}-3\right)\nonumber\\
  &&+\frac{2}{3} a_0 k_1^2 \left(-13 s^2+24 \left(\sqrt{s^2+1}-1\right)
  +\log \left(s^2+1\right)\right)\nonumber\\
  &&+\frac{2 a_0^2 k_1^2 \left(\left(24-11 \sqrt{s^2+1}\right) s^2
    -30 \left(\sqrt{s^2+1}-1\right)+2 \sqrt{s^2+1} \log
    \left(s^2+1\right)\right)}{3 \sqrt{s^2+1}},
\;\;\;\;\;\;\;
\end{eqnarray}
\begin{eqnarray}\label{4dBCFTCBChatQQ2} 
  \hat{Q}_2(s)&=&\frac{1}{3} \ha_2 \left(\left(2 \sqrt{s^2+1}-3\right) s^2
  +2 \left(\sqrt{s^2+1}-1\right)\right)\nonumber\\
  &&+\frac{4 k_1^2 \left(-4 s^4+s^2-5 \sqrt{s^2+1}+5\right)}
  {9 \sqrt{s^2+1}}\nonumber\\
  &&+\frac{2 a_0 k_1^2 \left(-26 s^4+20 s^2-46 \sqrt{s^2+1}
    +3 \sqrt{s^2+1} \log \left(s^2+1\right)+46\right)}
  {9 \sqrt{s^2+1}}\nonumber\\
  &&+\frac{4 a_0^2 k_1^2 \left(-11 s^4+14 s^2-34 \sqrt{s^2+1}
    +3 \sqrt{s^2+1} \log \left(s^2+1\right)+34\right)}{9 \sqrt{s^2+1}},
\end{eqnarray}
\begin{eqnarray}\label{4dBCFTCBChatQQ3} 
  \hat{Q}_3(s)&=&\frac{\left(3 \ha_1+2 \ha_2\right)
    \left(-2 s^4+\left(3 \sqrt{s^2+1}-4\right) s^2
    +2 \left(\sqrt{s^2+1}-1\right)\right)}{6 \sqrt{s^2+1}}\nonumber\\
  &&+\frac{4 k_1^2 \left(s^4+\left(6 \sqrt{s^2+1}+2\right) s^2
    -\sqrt{s^2+1}+1\right)}{9 \sqrt{s^2+1}}\nonumber\\
  &&+\frac{2 a_0^2 k_1^2 \left(46 s^4+\left(56-57 \sqrt{s^2+1}\right) s^2
    -10 \left(\sqrt{s^2+1}-1\right)+6 \sqrt{s^2+1}
    \log \left(s^2+1\right)\right)}{9 \sqrt{s^2+1}}\nonumber\\
  &&+\frac{2 a_0 k_1^2 \left(2 \left(s^2-3 \sqrt{s^2+1}+2\right) s^2
    -2 \sqrt{s^2+1}+3 \sqrt{s^2+1} \log \left(s^2+1\right)+2\right)}
      {9 \sqrt{s^2+1}},
\end{eqnarray}
where the integral constants are given by 
\begin{eqnarray}\label{4dBCFTCBCa1} 
  &&a_0=-\frac{\coth \rho }{(\coth \rho +1)^2},\\
  \label{4dBCFTCBCa2} 
  &&a_1=a_2=\frac{1}{48} k_1^2 e^{-4 \rho } (-8 \sinh 4 \rho 
  +\cosh 4 \rho -9),\\
  \label{4dBCFTCBCc1} 
  &&\ha_1=\frac{-k_1^2 e^{-3 \rho } (29 \sinh 3 \rho -
    54 \sinh \rho +19 \sinh 5 \rho +12 \cosh \rho 
    +22 \cosh 3 \rho -2 \cosh 5 \rho )}
        {18 \left(e^{2 \rho }+3\right)},\nonumber\\
        \label{4dBCFTCBCc3} 
&&\\
  &&\ha_2 =\frac{k_1^2 e^{-3 \rho } (4 \sinh \rho -4 \sinh 5 \rho 
    +8 \cosh \rho +3 \cosh 3 \rho +5 \cosh 5 \rho )}
  {2 \left(e^{2 \rho }+3\right)}.
\end{eqnarray}
Note that, according to \cite{Miao:2017aba}, suitable analytical
continuations should be performed for the above equations. In order to
get  continuous  solutions around  $x=0$, one should replace
$\sqrt{1+s^2}=\sqrt{1+\frac{z^2}{x^2}}$ by $\sqrt{1+s^2}=\sqrt{x^2+z^2}/x$
when $s=z/x$ is substituted
\cite{Miao:2017aba}.

The parameters of embedding function (\ref{Q1withb}) can be fixed by
the CBC (\ref{CBCa}),
\begin{eqnarray}
  d_1&=&\frac{1}{8} k_1^2 e^{-4 \rho } \sinh ^2 2 \rho  \sinh 4 \rho 
  \text{csch}\rho  \label{4dBCFTCBCd1} \\
  &&-\frac{1}{96} k_1^2 e^{-4 \rho } \sinh \rho  \left(8 \cosh ^2\rho 
  +(\cosh 2 \rho +4) \log \left(\coth ^2\rho \right)\right)\nonumber\\
  &&+\frac{1}{96} k_1^2 e^{-4 \rho } \sinh \rho  \left(4 \cosh 4 \rho 
  \left(\log \left(\coth ^2\rho \right)+5\right)+\cosh 6 \rho 
  \left(\log \left(\coth ^2\rho \right)+12\right)\right), \nn\\
  d_2&=&\frac{1}{24} k_1^2 e^{-4 \rho } (-5 \cosh 2 \rho +7 \cosh 4 \rho 
  -3 \cosh (6 \rho )+2) \label{4dBCFTCBCd2} \\
  &&\frac{1}{96} k_1^2 e^{-4 \rho } \sinh 4 \rho  \left(28-3 \sinh ^2 2 \rho 
  \log \left(\coth ^2 \rho \right)\right)\nonumber\\
  &&-\frac{1}{96} k_1^2 e^{-4 \rho } (33 \sinh 2 \rho +9 \sinh 6 \rho 
  +4 \sinh 8 \rho +4 \cosh 8 \rho ). \nn
\end{eqnarray}
We note that the coefficients $a_1, a_2, \ha_1, \ha_2$, $d_1$, $d_2$, and hence
the functions $X$, $\hat{X}$, $Q_{ab}$ and $\hat{Q}_{ab}$ are all
second order in the
extrinsic curvature $k_{ab}$. This is consistent with the assumption made in
the ansatz \eq{bulkmetricwithb}. 

\section{Mass spectrum for massive bulk scalar field}

Let us start with the EOM of the scalar field in the bulk
 \begin{eqnarray}\label{EOMscalar}
(\nabla_{\mu} \nabla^{\mu}-M^2) \phi(r,y)=0,
 \end{eqnarray}
 where $M$ is the mass of bulk scalar. Separating variables
 $\phi(r,y)=F(r) \bar{\phi} (y)$, we get
 \begin{eqnarray}\label{EOMmassivescalar}
&& \left(D_i D^i-m^2\right)\bar{\phi} (y)=0,\\
   && \cosh^2(r) F''(r)+d \sinh (r) \cosh (r) F'(r) +
   \left( m^2-M^2 \cosh ^2(r) \right) F(r)=0,
   \label{EOMmassiveF}
\end{eqnarray}
 Note that EOM of $F(r)$ is exactly the same as that of $H(r)$
 (\ref{EOMMBCmassiveH}) when $M=0$. In general, we
 have
 \begin{eqnarray}\label{massiveFsolution}
   F(r)=\text{sech}^{\frac{d}{2}}(r) \left(c_1 P_{\l}^{\mu}(\tanh r)
   +c_2 Q_{\l}^{\mu}(\tanh r)\right),
\end{eqnarray}
where $\lambda$ is given by (\ref{aibia1}) and 
 \begin{eqnarray}\label{Fmu}
\mu=\frac{\sqrt{d^2+4M^2}}{2}.
\end{eqnarray}
 From (\ref{massiveFsolution}), we get the BF bound that
 $M^2\ge -d^2/4$ and $m^2\ge -(d-1)^2/4$.

We impose DBC on the AdS boundary
 \begin{eqnarray}\label{DBCscalarM}
F(\infty)=0.
\end{eqnarray}
 On the end-of-world brane, we can impose either
 the CBC/DBC
 \begin{eqnarray}\label{DBCscalarQ}
F(-\rho)=0,
\end{eqnarray}
or the NBC
 \begin{eqnarray}\label{NBCscalarQ}
F'(-\rho)=0.
\end{eqnarray}

According to \cite{book}, in the large $r$ limit, we have
 \begin{eqnarray}\label{largerlimitQ}
   \lim_{r\to \infty}\text{sech}^{\frac{d}{2}}(r) Q_{\l}^{\mu}(\tanh r)
   \approx 2^{\frac{d+2\mu-4}{4}} \Gamma(\mu)\cos(\mu \pi)
   \left(1-\tanh(r)\right)^{\frac{d-\sqrt{d^2+4M^2}}{4}},
\end{eqnarray}
and
\begin{equation}\label{largerlimitP}
  \lim_{r\to \infty}\text{sech}^{\frac{d}{2}}(r) P_{\l}^{\mu}(\tanh r)
  \approx \begin{cases}
    \frac{2^{\frac{d+2\mu}{4}}}{\Gamma(1-\mu)}
   \left(1-\tanh(r)\right)^{\frac{d-\sqrt{d^2+4M^2}}{4}},&\
\mu \ \text{is non-integer} ,\\
c_3 \left(1-\tanh(r)\right)^{\frac{d+\sqrt{d^2+4M^2}}{4}},&\
\mu \ \text{is integer}.
\end{cases}
\end{equation}
where $c_3$ is a non-zero factor. 
As a result, we notice that, when
$M^2<0$, BC (\ref{DBCscalarM}) does not impose any constraint on the
solution (\ref{massiveFsolution})
and the spectrum of $m^2$
is continuous when $M^2<0$.  Since we have already studied the case
$M^2=0$ in sect.5, we focus on $M^2>0$ below.
We have two cases:

\text{Case 1:} When $M^2>0$ and $\mu$ is an integer, from (\ref{DBCscalarM}) we get
\begin{eqnarray}\label{massiveFcase1}
   F(r)=c_1 \text{sech}^{\frac{d}{2}}(r) P_{\l}^{\mu}(\tanh r).
\end{eqnarray}

\text{Case 2:} When $M^2>0$ and $\mu$ is
not an integer, from (\ref{DBCscalarM})  we derive
\begin{eqnarray}\label{massiveFcase3}
  F(r)=c_0 \ \text{sech}^{\frac{d}{2}}(r) \left( \Gamma(\mu) \cos(\mu \pi)
  P_{\l}^{\mu}(\tanh r)
   -\frac{2}{\Gamma(1-\mu) } Q_{\l}^{\mu}(\tanh r)\right),
\end{eqnarray}
where $c_0$ is a constant. 
Substituting (\ref{massiveFcase1},
\ref{massiveFcase3})
into the BCs (\ref{DBCscalarQ},\ref{NBCscalarQ}) on $Q$, we get the
constraint of mass spectrum for various cases discussed above.

\section{Localization of gravity}

In this appendix, we discuss the localization of gravity in
the brane
world theory \cite{Randall:1999ee,Randall:1999vf,Karch:2000ct}. For
simplicity, we focus on the AdS brane. We find that, for NBC, the
graviton is located on the brane with non-negative tension, while for
CBC/DBC, the graviton is located on the brane with non-positive
tension.  For CBC with zero brane tension, we verify that the first
mode of graviton is located on the brane in the sense that the wave
function peaks on the brane only and decays when it goes far from the
brane.

Before we start, it should be stressed that although the existence of
localized gravity is important for brane world theory with a two-side
brane \cite{Randall:1999ee,Randall:1999vf,Karch:2000ct}, it is not a
necessary condition for AdS/BCFT with a one-side end-of-the-world
brane \cite{Takayanagi:2011zk}. That is because brane world theory and
AdS/BCFT have different physical motivations. On one hand, brane world
theory aims to describe our real world. To do so, one requires that
the gravity is localized on the brane and the gravity mass is
sufficiently small, so that general relativity can be recovered at low
energies.  On the other hand, AdS/BCFT is a gravity dual of BCFT. It
is well-defined as long as it gives the reasonable BCFT data.

\subsection{Volcanic potential}
\label{vp}

To analyze the localization of gravity, let us study the ``volcanic
potential".  To recover the delta function in the potential, we
rewrite perturbation metric (\ref{perturbationmetric}) as
 \begin{eqnarray}\label{perturbationmetric1}
ds^2=dr^2+\exp(2A(r)) \left( \bar{h}^{(0)}_{ij}(y) + \epsilon H(r)
   \bar{h}^{(1)}_{ij}(y)  \right)dy^i dy^j+O(\epsilon^2),
\end{eqnarray}
 where the brane is located at $r=-\rho$ and $\exp(2A(r))=
 \cosh^2(-\rho+|r+\rho|)$. 
Now the equation of motion (EOM) of $H(r)$ (\ref{EOMMBCmassiveH}) becomes
\begin{eqnarray}\exp\left(2A(r)\right) \Big( H''(r)+d  A'(r)H'(r) \Big)
  + m^2 H(r)=0.
   \label{EOMMBCmassiveH1}
\end{eqnarray}

To warm-up, we first study the case of NBC, where the dynamical
gravity on the brane is given by the induced metric
(\ref{dynamicalgravity}), which is proportional to $H(r)$. Performing
the transformations
  \begin{eqnarray}\label{transformationsNBC}
    dr=\exp\left(A(w)\right) dw, \  H(r)=\exp\left(-\frac{d-1}{2}A(w)\right)
    \Psi_{\text{NBC}}(w),
 \end{eqnarray}
 we rewrite (\ref{EOMMBCmassiveH1}) as 
  \begin{eqnarray}\label{EOMII}
-\Psi_{\text{NBC}}''+ V_{\text{NBC}}  \Psi_{\text{NBC}}=m^2 \Psi_{\text{NBC}},
  \end{eqnarray}
where the potential energy is given by
\begin{eqnarray}\label{VNBC1}
V_{\text{NBC}}=\frac{1}{4} (d-1) \Big(2 A''(w)+(d-1) A'(w)^2\Big).
  \end{eqnarray}
Note that the delta function is hidden in $ A''(w)$.
From $\exp(A(r))=\cosh(-\rho+|r+\rho|)$ and (\ref{transformationsNBC}), we have
\begin{eqnarray}\label{transformationsNBC1}
A(w)=-\log \cos(w_0+|w-w_0|),  \ \ \  w_0=-\sin^{-1}(\text{tanh}(\rho)).
\end{eqnarray}
Substituting the above equations into (\ref{VNBC1}),
we derive the volcanic potential for NBC
\begin{eqnarray}\label{VNBC}
  V_{\text{NBC}}=\frac{1}{4} (d-1)
  \Big( (d+1) \sec ^2(w_0+|w-w_0|)-d+1\Big)-(d-1)\sinh (\rho )
 \delta(w-w_0),
 \end{eqnarray}
 which agrees with \cite{Karch:2000ct}.

Now let us turn to the case of CBC/DBC. Recall that, for CBC/DBC, the
dynamical gravity on the brane is given by the extrinsic curvature
(\ref{dynamicalgravity}), which is proportional to $H'(r)$.
Differentiating (\ref{EOMMBCmassiveH}), we get EOM of $H'(r)$
 \begin{eqnarray}\label{EOMCBC}
   e^{2 A(r)} \left(H^{(3)}(r)+(d+2) A'(r) H''(r)\right)
   +\left(d e^{2 A(r)} A''(r)+2 d e^{2 A(r)} A'(r)^2+m^2\right)H'(r) =0.
 \end{eqnarray}
 Performing the transformations
  \begin{eqnarray}\label{transformationsCBC}
    dr=\exp\left(A(w)\right) dw, \  H'(r)=\exp\left(-\frac{d+1}{2}A(w)\right)
    \Psi_{\text{CBC}}(w),
 \end{eqnarray}
 we rewrite (\ref{EOMCBC}) as 
  \begin{eqnarray}\label{EOMCBC1}
-\Psi_{\text{CBC}}''+ V_{\text{CBC}}  \Psi_{\text{CBC}}=m^2 \Psi_{\text{CBC}}
  \end{eqnarray}
  where 
  \begin{eqnarray}\label{VCBC1}
V_{\text{CBC}}=\frac{1}{4} (d-1) \Big(-2 A''(w)+(d-1) A'(w)^2\Big).
  \end{eqnarray}
  By applying (\ref{transformationsNBC1}), we finally derive the
  potential for CBC
  \begin{eqnarray}\label{VCBC}
    V_{\text{CBC}}=\frac{1}{4} (d-1) \Big((d-3)
    \sec ^2(w_0+|w-w_0|)-d+1\Big)+(d-1)\sinh (\rho )
 \delta(w-w_0).
  \end{eqnarray}
  
  Note that the delta function has opposite sign in the potential of
  NBC (\ref{VNBC}) and CBC (\ref{VCBC}).  At low energies, the gravity
  tends to be located at the minimum of the potential energy. Thus we
  hope to have a negative delta function potential in order to have a
  located gravity on the brane. This is the case for NBC with a
  positive brane tension $T=(d-1)\tanh(\rho)>0$, while for CBC with a
  negative brane tension $T=(d-1)\tanh(\rho)<0$.  As a result, for
  CBC, the gravity is well
  localized on the brane with a negative
  tension instead of a positive tension.  It is interesting to study
  the critical case of a zero tension.
  For zero brane tension, we have $w_0=\rho=0$. As a
  result, for $d\ge 4$ the potential $V_{\text{CBC}}$ (\ref{VCBC}) is
  minimum at $w=w_0=0$ and goes to infinity faraway from the
  brane. Thus it is expected that the gravity is located on the brane
  with zero tension at low energies. As we will show in the next
  subsection, this is indeed the case.

\subsection{Wave function}

In this subsection, we show that, for CBC with zero brane tension, the
first mode of graviton is located on the brane in the sense that the
wave function peaks on the brane only and decays when it goes far from
the brane.  For simplicity, let us focus on four dimensions $d=4$ and
only one side of the brane $r \ge 0$, the other side with $r\le 0$ can
be recovered by the $Z_2$ symmetry.

For phenomenological reasons, we hope the mass of first mode is
small.
This can be achieved with a two branes setup as follow: One brane is
set at $r=0$
in the bulk, and the other one is located at $r=\tanh ^{-1}(\cos
(\epsilon ))$
near the AdS boundary.
Here  $\epsilon \neq 0$ is a small parameter.
We impose CBC on the
brane in the bulk, while NBC on the brane close to the AdS boundary
 \begin{eqnarray}\label{CBCNBC}
H(0)=0,\quad  H'(\tanh ^{-1}(\cos (\epsilon )))=0.
 \end{eqnarray}
We will show now that the smaller $\epsilon$ is, the
closer the second brane approaches to the AdS boundary, and the
smaller the mass of first mode is.
In fact, substituting the general solution (\ref{massiveHsolution}) of $H(r)$ 
into the BCs (\ref{CBCNBC}), we obtain a constraint equation for the
mass. Focus on the first mode, we have
\begin{eqnarray}\label{massMBC1}
&&m^2=0.02969, \ \ \text{for}\ \epsilon=0.1,\\
&&m^2=0.00029, \ \ \text{for} \ \epsilon=0.01, \label{massMBC2}
 \end{eqnarray}
which shows 
it is roughly that $m^2 \propto \epsilon^2$ for the first mode of graviton. 
We require that the first mode is normalizable
\begin{eqnarray}\label{normalizable}
  1=\int_0^{\frac{\pi}{2}-\epsilon} \Psi_{\text{CBC}}^2(w) dw=
  \int_0^{\tanh ^{-1}(\cos (\epsilon ))} \cosh^{d}(r) H'^2(r) dr,
 \end{eqnarray}
 where 
  \begin{eqnarray}\label{wavefucntionCBC}
\Psi_{\text{CBC}}(w)=\cosh^{\frac{d+1}{2}}(r) H'(r),
 \end{eqnarray}
is the wave function (\ref{transformationsCBC}) for CBC/DBC.
 Eq.(\ref{normalizable}) together with BCs (\ref{CBCNBC}) can fix the
 integral constants of $H(r)$ (\ref{massiveHsolution}) numerically.
 Note that, one cannot set $\epsilon=0$ in order to have a
 normalizable wave function. This means that the gravitons on the
 brane are always massive.

\begin{figure}[ht!]
\centering
\includegraphics[width=7cm]{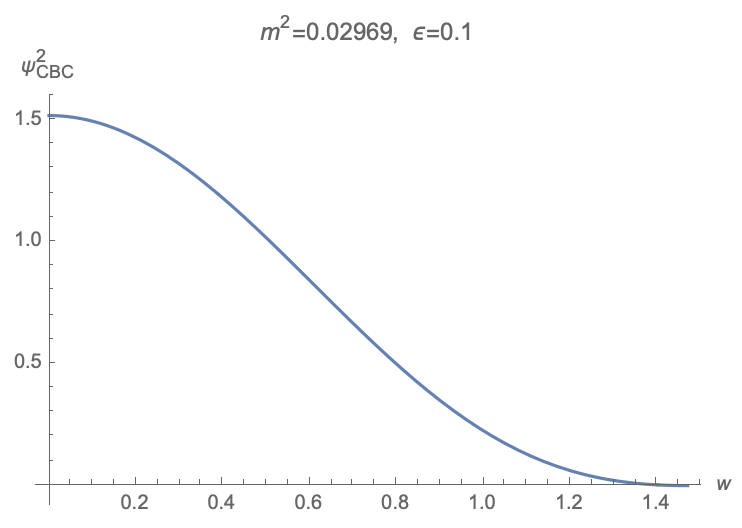}
\includegraphics[width=7cm]{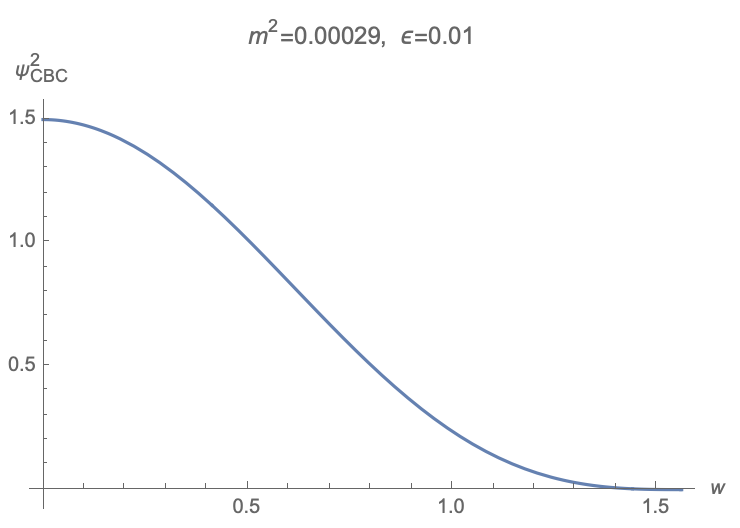}
\caption{For CBC with zero brane tension, the first mode of massive
  gravity is localized on the brane in the sense that
  $\Psi^2_{\text{CBC}}(w)$ peaks on the brane $w = 0$ only and decays
  when it goes far from the brane.}
\label{locatedgravityCBC}
\end{figure}

\begin{figure}[ht!]
\centering
\includegraphics[width=7cm]{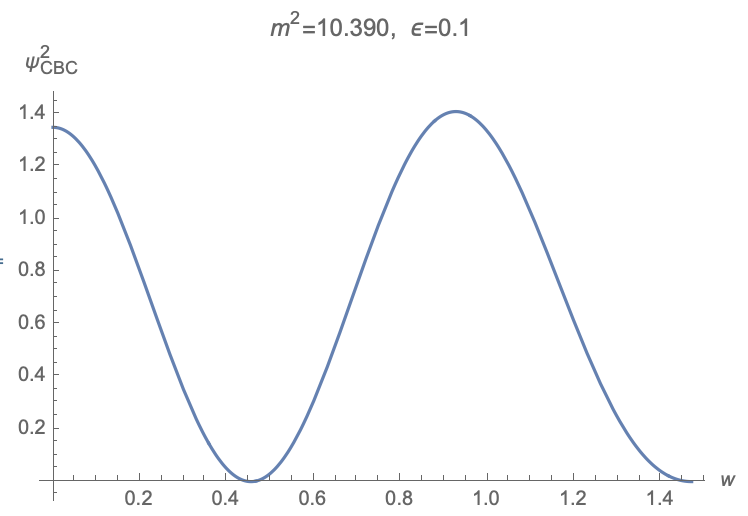}
\includegraphics[width=7cm]{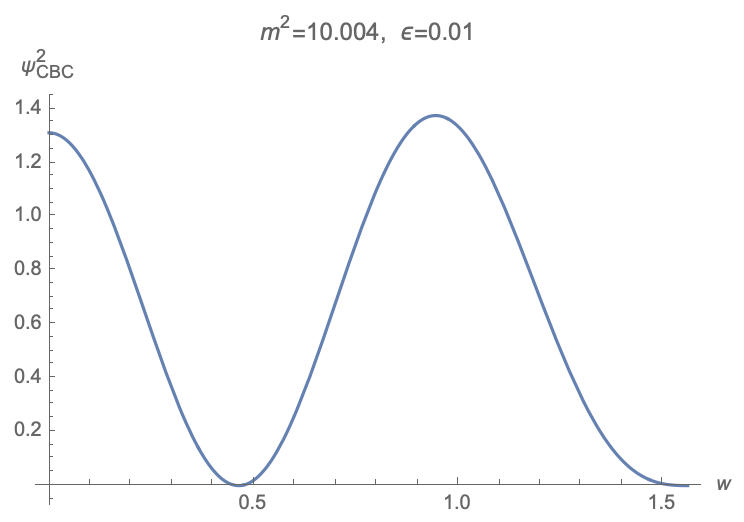}
\caption{For CBC with zero brane tension, the second mode of massive
  gravity is not well-localized on the brane $w=0$.}
\label{secondmodeCBC}
\end{figure}

 To see the localization of gravity, let us draw the figure of the
 wave function $\Psi_{\text{CBC}}(w)$ for CBC.  As shown in
 Fig. \ref{locatedgravityCBC}, the first mode of massive gravity is
 localized on the brane $w=0$ in the sense that
 $\Psi_{\text{CBC}}^2(w)$ peaks at $w
 = 0$ only and decays when it goes far from the brane. On the other
 hand, similar to the case of NBC, the other modes are not
 well-localized on the brane
as shown by the oscillation behaviour of the wave function.
 See Fig. \ref{secondmodeCBC}.

\begin{figure}[ht!]    
\centering
\includegraphics[width=7cm]{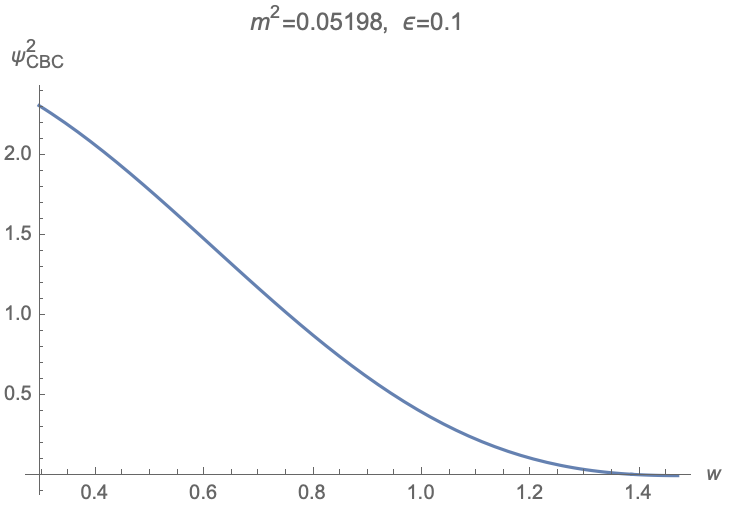}
\includegraphics[width=7cm]{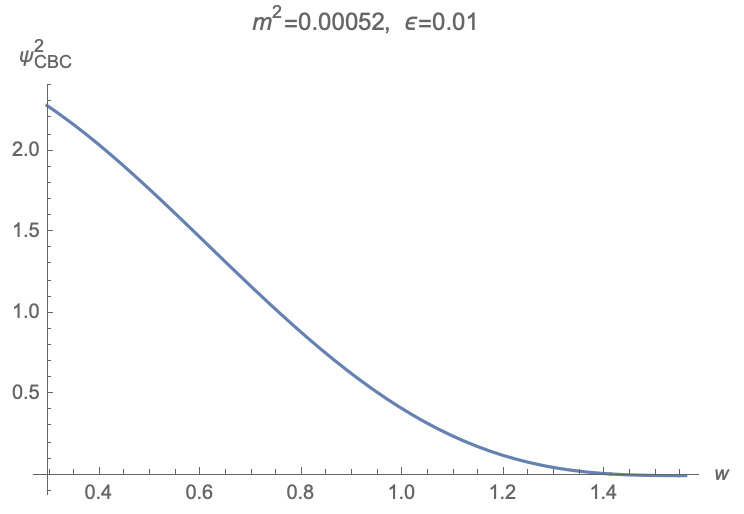}
\caption{For CBC with negative brane tension $T=3 \tanh\rho\approx
  -0.874$, the first mode of gravity is localized on the brane in the
  sense that $\Psi^2_{\text{CBC}}(w)$ peaks on the brane $w\approx
  0.296$ only and decays when it goes far from the brane.}
\label{negativeTCBC}
\end{figure}

\begin{figure}[ht!]    
\centering
\includegraphics[width=7cm]{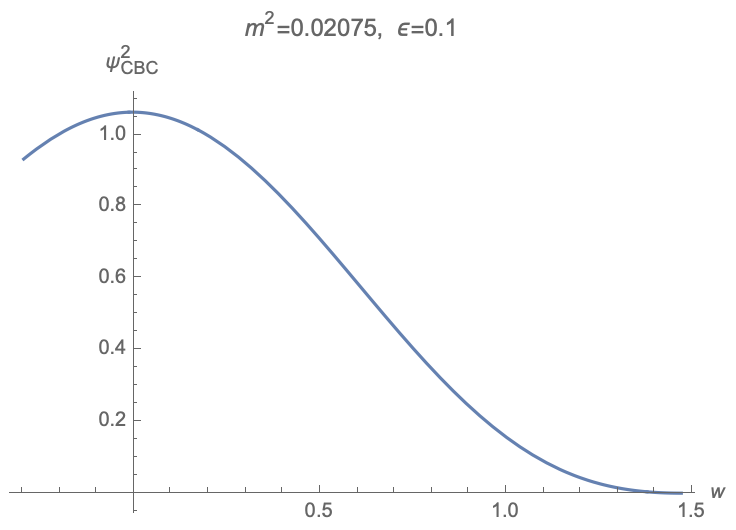}
\includegraphics[width=7cm]{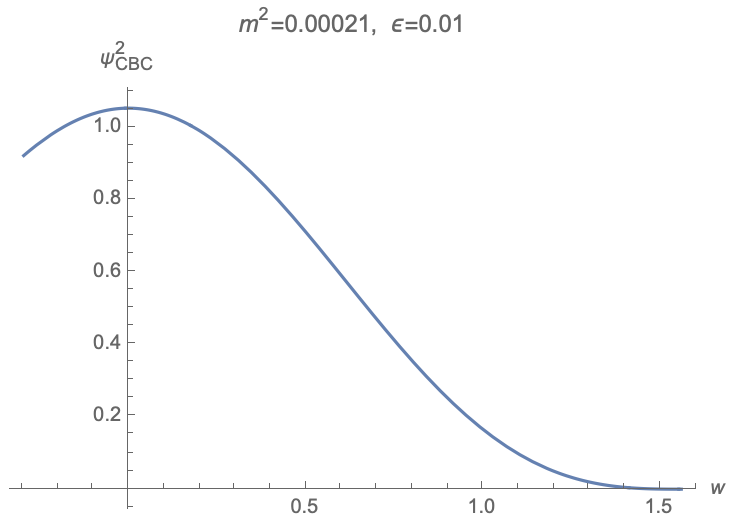}
\caption{For CBC with positive brane tension $T=3 \tanh\rho\approx
  0.874$, the first mode of gravity is not well-localized on the brane
  at $w\approx-0.296$.}
\label{positiveTCBC}
\end{figure}

In the above discussions, we
have focused on the zero brane tension.  To end
 this subsection, let us briefly discuss the case of non-zero brane
 tension.  Following the above approach, let us draw the figures of
 wave functions.  As shown in Fig. {\ref{negativeTCBC}}, for CBC with
 negative brane tension, the first mode of gravity is localized on the
 brane. While for CBC with positive brane tension, as shown in
 Fig. {\ref{positiveTCBC}}, the gravity is not well-localized on the
 brane.  This agrees with the analysis of ``volcanic potential'' in
 section
 \ref{vp}.


\begin{thebibliography}{00}


\bibitem{Witten:2018lgb} 
  E.~Witten,
  arXiv:1805.11559 [hep-th].
  
\bibitem{Miao:2017gyt}
R.~X.~Miao, C.~S.~Chu and W.~Z.~Guo,
Phys. Rev. D \textbf{96}, no.4, 046005 (2017)
[arXiv:1701.04275 [hep-th]].

\bibitem{Chu:2017aab}
C.~S.~Chu, R.~X.~Miao and W.~Z.~Guo,
JHEP \textbf{04}, 089 (2017)
[arXiv:1701.07202 [hep-th]].

\bibitem{Penington:2019npb}
G.~Penington,
JHEP \textbf{09}, 002 (2020)
[arXiv:1905.08255 [hep-th]].

\bibitem{Almheiri:2019psf}
A.~Almheiri, N.~Engelhardt, D.~Marolf and H.~Maxfield,
JHEP \textbf{12}, 063 (2019)
[arXiv:1905.08762 [hep-th]].

\bibitem{Almheiri:2019hni}
A.~Almheiri, R.~Mahajan, J.~Maldacena and Y.~Zhao,
JHEP \textbf{03}, 149 (2020)
[arXiv:1908.10996 [hep-th]].


\bibitem{Rozali:2019day}
M.~Rozali, J.~Sully, M.~Van Raamsdonk, C.~Waddell and D.~Wakeham,
JHEP \textbf{05}, 004 (2020)
[arXiv:1910.12836 [hep-th]].

\bibitem{Chen:2019uhq}
H.~Z.~Chen, Z.~Fisher, J.~Hernandez, R.~C.~Myers and S.~M.~Ruan,
JHEP \textbf{03}, 152 (2020)
[arXiv:1911.03402 [hep-th]].

\bibitem{Almheiri:2019psy}
A.~Almheiri, R.~Mahajan and J.~E.~Santos,
SciPost Phys. \textbf{9}, no.1, 001 (2020)
[arXiv:1911.09666 [hep-th]].

\bibitem{Kusuki:2019hcg}
Y.~Kusuki, Y.~Suzuki, T.~Takayanagi and K.~Umemoto,
[arXiv:1912.08423 [hep-th]].

\bibitem{Balasubramanian:2020hfs}
V.~Balasubramanian, A.~Kar, O.~Parrikar, G.~Sárosi and T.~Ugajin,
[arXiv:2003.05448 [hep-th]].

\bibitem{Sully:2020pza}
J.~Sully, M.~Van Raamsdonk and D.~Wakeham,
[arXiv:2004.13088 [hep-th]].


\bibitem{Geng:2020qvw}
H.~Geng and A.~Karch,
JHEP \textbf{09} (2020), 121
[arXiv:2006.02438 [hep-th]].

\bibitem{Chen:2020uac}
H.~Z.~Chen, R.~C.~Myers, D.~Neuenfeld, I.~A.~Reyes and J.~Sandor,
[arXiv:2006.04851 [hep-th]].

\bibitem{Dong:2020uxp}
X.~Dong, X.~L.~Qi, Z.~Shangnan and Z.~Yang,
[arXiv:2007.02987 [hep-th]].

\bibitem{Arias:2019zug}
C.~Arias, F.~Diaz, R.~Olea and P.~Sundell,
JHEP \textbf{04}, 124 (2020)
[arXiv:1906.05310 [hep-th]].

\bibitem{Arias:2019pzy}
C.~Arias, F.~Diaz and P.~Sundell,
Class. Quant. Grav. \textbf{37}, no.1, 015009 (2020)
[arXiv:1901.04554 [hep-th]].

\bibitem{Geng:2020kxh}
H.~Geng,
[arXiv:2005.00021 [hep-th]].

\bibitem{Ling:2020laa}
Y.~Ling, Y.~Liu and Z.~Y.~Xian,
[arXiv:2010.00037 [hep-th]].

\bibitem{Geng:2020fxl}
H.~Geng, A.~Karch, C.~Perez-Pardavila, S.~Raju, L.~Randall, M.~Riojas and S.~Shashi,
[arXiv:2012.04671 [hep-th]].

\bibitem{Kawabata:2021hac}
K.~Kawabata, T.~Nishioka, Y.~Okuyama and K.~Watanabe,
JHEP \textbf{05}, 062 (2021)
[arXiv:2102.02425 [hep-th]].

\bibitem{Bhattacharya:2021jrn}
A.~Bhattacharya, A.~Bhattacharyya, P.~Nandy and A.~K.~Patra,
JHEP \textbf{05}, 135 (2021)
[arXiv:2103.15852 [hep-th]].

\bibitem{Kawabata:2021vyo}
K.~Kawabata, T.~Nishioka, Y.~Okuyama and K.~Watanabe,
[arXiv:2105.08396 [hep-th]].

\bibitem{Geng:2021hlu}
H.~Geng, A.~Karch, C.~Perez-Pardavila, S.~Raju, L.~Randall, M.~Riojas and S.~Shashi,
[arXiv:2107.03390 [hep-th]].

\bibitem{Krishnan:2020fer}
C.~Krishnan,
JHEP \textbf{01}, 179 (2021)
[arXiv:2007.06551 [hep-th]].

\bibitem{Deng:2020ent}
F.~Deng, J.~Chu and Y.~Zhou,
JHEP \textbf{03}, 008 (2021)
[arXiv:2012.07612 [hep-th]].

\bibitem{Chu:2021gdb}
J.~Chu, F.~Deng and Y.~Zhou,
[arXiv:2105.09106 [hep-th]].

\bibitem{Neuenfeld:2021wbl}
D.~Neuenfeld,
[arXiv:2104.02801 [hep-th]].

\bibitem{Neuenfeld:2021bsb}
D.~Neuenfeld,
[arXiv:2105.01130 [hep-th]].

\bibitem{Chen:2020hmv}
H.~Z.~Chen, R.~C.~Myers, D.~Neuenfeld, I.~A.~Reyes and J.~Sandor,
JHEP \textbf{12}, 025 (2020)
[arXiv:2010.00018 [hep-th]].

\bibitem{Ghosh:2021axl}
K.~Ghosh and C.~Krishnan,
JHEP \textbf{08}, 119 (2021)
[arXiv:2103.17253 [hep-th]].

\bibitem{Omiya:2021olc}
H.~Omiya and Z.~Wei,
[arXiv:2107.01219 [hep-th]].



\bibitem{Maldacena:1997re}
  J.~M.~Maldacena,
  Int.\ J.\ Theor.\ Phys.\  {\bf 38}, 1113 (1999)
  [Adv.\ Theor.\ Math.\ Phys.\  {\bf 2}, 231 (1998)]
  [hep-th/9711200].
  
   
\bibitem{Gubser:1998bc}
S.~S.~Gubser, I.~R.~Klebanov and A.~M.~Polyakov,
Phys. Lett. B \textbf{428}, 105-114 (1998)
[arXiv:hep-th/9802109 [hep-th]].

\bibitem{Witten:1998qj}
E.~Witten,
Adv. Theor. Math. Phys. \textbf{2}, 253-291 (1998)
[arXiv:hep-th/9802150 [hep-th]].






\bibitem{Randall:1999ee}
L.~Randall and R.~Sundrum,
Phys. Rev. Lett. \textbf{83}, 3370-3373 (1999)
[arXiv:hep-ph/9905221 [hep-ph]].

\bibitem{Randall:1999vf}
L.~Randall and R.~Sundrum,
Phys. Rev. Lett. \textbf{83}, 4690-4693 (1999)
[arXiv:hep-th/9906064 [hep-th]].

\bibitem{Karch:2000ct}
A.~Karch and L.~Randall,
JHEP \textbf{05}, 008 (2001)
[arXiv:hep-th/0011156 [hep-th]].


\bibitem{Takayanagi:2011zk}
  T.~Takayanagi,
  Phys.\ Rev.\ Lett.\  {\bf 107} (2011) 101602
  [arXiv:1105.5165 [hep-th]].
  
\bibitem{Fujita:2011fp}
M.~Fujita, T.~Takayanagi and E.~Tonni,
JHEP \textbf{11}, 043 (2011)
[arXiv:1108.5152 [hep-th]].

\bibitem{Nozaki:2012qd}
M.~Nozaki, T.~Takayanagi and T.~Ugajin,
JHEP \textbf{06}, 066 (2012)
[arXiv:1205.1573 [hep-th]].

\bibitem{Miao:2018qkc}
R.~X.~Miao,
JHEP \textbf{02}, 025 (2019)
[arXiv:1806.10777 [hep-th]].


\bibitem{Cardy:2004hm}
  J.~L.~Cardy,
  hep-th/0411189.

\bibitem{McAvity:1993ue} 
  D.~M.~McAvity and H.~Osborn,
  Nucl.\ Phys.\ B {\bf 406}, 655 (1993)
  [hep-th/9302068].


\bibitem{Akal:2020wfl}
I.~Akal, Y.~Kusuki, T.~Takayanagi and Z.~Wei,
Phys. Rev. D \textbf{102}, no.12, 126007 (2020)
[arXiv:2007.06800 [hep-th]].

\bibitem{Bousso:2020kmy}
R.~Bousso and E.~Wildenhain,
Phys. Rev. D \textbf{102}, no.6, 066005 (2020)
[arXiv:2006.16289 [hep-th]].

\bibitem{Miao:2020oey}
R.~X.~Miao,
JHEP \textbf{01}, 150 (2021)
[arXiv:2009.06263 [hep-th]].



\bibitem{Miao:2021ual}
R.~X.~Miao,
Phys. Rev. D \textbf{104}, no.8, 8 (2021)
[arXiv:2101.10031 [hep-th]].

\bibitem{Miao:2017aba} 
  R.~X.~Miao and C.~S.~Chu,
  JHEP {\bf 1803}, 046 (2018)
  [arXiv:1706.09652 [hep-th]].
  


\bibitem{Anderson:2006lqb}
M.~T.~Anderson,
Geom. Topol. \textbf{12}, no.4, 2009-2045 (2008)
[arXiv:math/0612647 [math.DG]].

\bibitem{Anderson:2007jpe}
M.~T.~Anderson,
Selecta Math. 16 (2010) 343-375,
[arXiv:0704.3373 [math.DG]].

\bibitem{Anderson:2010ph}
M.~T.~Anderson,
Phys. Rev. D \textbf{82}, 084044 (2010)
[arXiv:1008.4309 [gr-qc]].

\bibitem{York:1972sj}
J.~W.~York, Jr.,
Phys. Rev. Lett. \textbf{28}, 1082-1085 (1972)

\bibitem{Papadimitriou:2005ii}
I.~Papadimitriou and K.~Skenderis,
JHEP \textbf{08}, 004 (2005)
[arXiv:hep-th/0505190 [hep-th]].

\bibitem{Deutsch:1978sc} 
  D.~Deutsch and P.~Candelas,
  Phys.\ Rev.\ D {\bf 20}, 3063 (1979).
  
\bibitem{Jensen:2015swa} 
  K.~Jensen and A.~O'Bannon,
  Phys.\ Rev.\ Lett.\  {\bf 116}, no. 9, 091601 (2016)
  [arXiv:1509.02160 [hep-th]].
   
\bibitem{Herzog:2015ioa} 
  C.~P.~Herzog, K.~W.~Huang and K.~Jensen,
  JHEP {\bf 1601}, 162 (2016)
  [arXiv:1510.00021 [hep-th]].
   
\bibitem{Fursaev:2015wpa} 
  D.~Fursaev,
  JHEP {\bf 1512}, 112 (2015)
 [arXiv:1510.01427 [hep-th]].

\bibitem{Solodukhin:2015eca} 
  S.~N.~Solodukhin,
  Phys.\ Lett.\ B {\bf 752}, 131 (2016)
[arXiv:1510.04566 [hep-th]].
  
  
\bibitem{Miao:2018dvm}
R.~X.~Miao,
JHEP \textbf{07}, 098 (2019)
[arXiv:1808.05783 [hep-th]].
   

\bibitem{Chu:2018ksb}
C.~S.~Chu and R.~X.~Miao,
Phys. Rev. Lett. \textbf{121}, no.25, 251602 (2018)
[arXiv:1803.03068 [hep-th]].

\bibitem{Chu:2018ntx}
C.~S.~Chu and R.~X.~Miao,
JHEP \textbf{07}, 005 (2018)
[arXiv:1804.01648 [hep-th]].

\bibitem{Chu:2018fpx}
C.~S.~Chu and R.~X.~Miao,
JHEP \textbf{07}, 151 (2019)
[arXiv:1812.10273 [hep-th]].

\bibitem{Chu:2019rod}
C.~S.~Chu,
Fortsch. Phys. \textbf{67}, no.8-9, 1910005 (2019)
[arXiv:1903.02817 [hep-th]].

\bibitem{Chu:2020mwx}
C.~S.~Chu and R.~X.~Miao,
Phys. Rev. D \textbf{102}, no.4, 046011 (2020)
[arXiv:2004.05780 [hep-th]].

\bibitem{Chu:2020gwq}
C.~S.~Chu and R.~X.~Miao,
JHEP \textbf{08}, 134 (2020)
[arXiv:2005.12975 [hep-th]].

\bibitem{Billo:2016cpy} 
  M.~Billò,  V .~Goncalves, E.~Lauria and M.~Meineri,
  JHEP {\bf 1604}, 091 (2016)
  [arXiv:1601.02883 [hep-th]].

 \bibitem{Herzog:2017kkj} 
  C.~Herzog, K.~W.~Huang and K.~Jensen,
  Phys.\ Rev.\ Lett.\  {\bf 120}, no. 2, 021601 (2018)
  [arXiv:1709.07431 [hep-th]].
  
\bibitem{Herzog:2017xha} 
  C.~P.~Herzog and K.~W.~Huang,
  JHEP {\bf 1710}, 189 (2017)
  [arXiv:1707.06224 [hep-th]].

\bibitem{Henningson:1998gx}
  M.~Henningson and K.~Skenderis,
  JHEP {\bf 9807} (1998) 023
  [hep-th/9806087].


 \bibitem{book} 
 Magnus W ,  Oberhettinger F ,  Soni R, ``Formulas and Theorems for the Special Functions of Mathematical Physics", 
 American Journal of Physics 35, 550 (1967).

\bibitem{Karch:2018uft}
A.~Karch and Y.~Sato,
JHEP \textbf{07} (2018), 156
[arXiv:1805.10427 [hep-th]].


\bibitem{Bissi:2018mcq}
A.~Bissi, T.~Hansen and A.~S\"oderberg,
JHEP \textbf{01}, 010 (2019)
[arXiv:1808.08155 [hep-th]].

\bibitem{Mazac:2018biw}
D.~Maz\'a\v{c}, L.~Rastelli and X.~Zhou,
JHEP \textbf{12}, 004 (2019)
[arXiv:1812.09314 [hep-th]].

\bibitem{Reeves:2021sab}
W.~Reeves, M.~Rozali, P.~Simidzija, J.~Sully, C.~Waddell and D.~Wakeham,
[arXiv:2108.10345 [hep-th]].

\bibitem{Kastikainen:2021ybu}
J.~Kastikainen and S.~Shashi,
[arXiv:2109.00079 [hep-th]].

\bibitem{Rychkov:2016iqz}
S.~Rychkov,
[arXiv:1601.05000 [hep-th]].

\bibitem{Simmons-Duffin:2016gjk}
D.~Simmons-Duffin,
[arXiv:1602.07982 [hep-th]].

\bibitem{Poland:2018epd}
D.~Poland, S.~Rychkov and A.~Vichi,
Rev. Mod. Phys. \textbf{91}, 015002 (2019)
doi:10.1103/RevModPhys.91.015002
[arXiv:1805.04405 [hep-th]].


\end{thebibliography}
\end{document}